\documentclass[final]{siamart1116}
\pdfoutput=1

\usepackage[utf8]{inputenc}
\usepackage[english]{babel}
\usepackage[noend]{algpseudocode}
\usepackage{amssymb}
\usepackage{caption}
\usepackage{subcaption}
\usepackage[inline]{enumitem}
\usepackage{listings}
\usepackage{color,soul}
\usepackage{multicol}
\usepackage{tikz}
\usepackage{verbatim}

\usetikzlibrary{arrows}
\usetikzlibrary{shapes}

\newcommand{\ds}{\, \mathrm{d}s}
\newcommand{\dx}{\, \mathrm{d}x}
\newcommand{\dX}{\, \mathrm{d}X}

\newcommand{\average}[1]{\ensuremath{\{#1\}}}
\newcommand{\jump}[1]{\ensuremath{[\![#1]\!]}}

\floatstyle{ruled}
\newfloat{listing}{thbp}{lst}
\floatname{listing}{Listing}

\numberwithin{equation}{section}

\newcommand{\TheTitle}{TSFC: a structure-preserving form compiler}
\newcommand{\TheAuthors}{M.~Homolya, L.~Mitchell, F.~Luporini, and D.~A.~Ham}

\headers{\TheTitle}{\TheAuthors}

\title{{\TheTitle}%
  \thanks{Submitted to SIAM Journal on Scientific Computing on \today.
\funding{This work was supported by The Grantham Institute; the
  Engineering and Physical Sciences Research Council [grant number
  EP/M011054/1]; the Department of Computing, Imperial College London;
  and the Natural Environment Research Council [grant number
  NE/K008951/1].}}}
\author{Miklós Homolya%
  \thanks{The Grantham Institute and Department of Computing, Imperial
    College London, London, SW7~2AZ, UK
    (\email{m.homolya14@imperial.ac.uk})}
  \and Lawrence Mitchell%
  \thanks{Department of Computing and Department of Mathematics,
    Imperial College London, London, SW7~2AZ, UK
    (\email{lawrence.mitchell@imperial.ac.uk})}
  \and Fabio Luporini%
  \thanks{Now at Department of Earth Science and Engineering, Imperial
    College London, London, SW7~2AZ, UK
    (\email{f.luporini12@imperial.ac.uk})}
  \and David A.\@ Ham%
  \thanks{Department of Mathematics, Imperial College London, London,
    SW7~2AZ, UK (\email{david.ham@imperial.ac.uk})}
}

\begin{document}

\maketitle

\begin{abstract}
  A \emph{form compiler} takes a high-level description of the weak
  form of partial differential equations and produces low-level code
  that carries out the finite element assembly.  In this paper we
  present the \emph{Two-Stage Form Compiler} (TSFC), a new form
  compiler with the main motivation to maintain the structure of the
  input expression as long as possible.  This facilitates the
  application of optimizations at the highest possible level of
  abstraction.  TSFC features a novel, structure-preserving method for
  separating the contributions of a form to the subblocks of the local
  tensor in discontinuous Galerkin problems.  This enables us to
  preserve the tensor structure of expressions longer through the
  compilation process than other form compilers.  This is also
  achieved in part by a two-stage approach that cleanly separates the
  lowering of finite element constructs to tensor algebra in the first
  stage, from the scheduling of those tensor operations in the second
  stage.  TSFC also efficiently traverses complicated expressions, and
  experimental evaluation demonstrates good compile-time performance
  even for highly complex forms.
\end{abstract}

\begin{keywords}
  code generation,
  finite element method,
  form compiler,
  tensor algebra,
  weak form
\end{keywords}

\begin{AMS}
  68N20,
  65M60,
  65N30
\end{AMS}

\section{Introduction}
\label{sec:intro}

The development of state-of-the-art finite element simulation software
has become an especially demanding endeavour.  Advanced finite element
techniques are particularly complicated, the efficient utilization of
modern computers requires onerous effort, and the application domain
itself is often a source of remarkable complexity.  Automatic code
generation techniques have been employed by various software projects
to decouple and manage these complexities. Examples include Analysa
\cite{Bagheri}, GetDP \cite{Dular1998}, Sundance \cite{Long2010},
FreeFem++ \cite{Hecht2012}, the FEniCS Project
\cite{logg2012automated,Alnaes2015}, and Firedrake
\cite{Rathgeber2016}. The finite element method is particularly suited to
code generation approaches since the entire mathematical
specification of a finite element problem can be expressed at a high level
of abstraction. The specification of the weak form of the
equations, along with the discrete function space from which each variable
is to be drawn, is sufficient to characterise the problem completely. The
\emph{Unified Form Language} (UFL) \cite{Alnes2014} is a domain
specific language for
the finite element method, and provides the input language for FEniCS and
Firedrake. UFL captures the weak form and discretised function spaces,
thereby providing the information that the code generation system requires. 

The solution of a finite element problem requires three distinct
operations. First, each integral in the problem (that is, each multilinear
form) must be evaluated on each cell or facet of the mesh. Second, the facet
and cell components are assembled into global matrices and/or vectors as
appropriate. Finally, these global objects are used by a linear solver to
compute the answer to the system. Time varying and nonlinear PDEs are solved
by composing these operations under the control of a time integration scheme
and nonlinear solver respectively.

A \emph{form compiler} takes a multilinear form expressed in UFL and
produces low-level code that assembles the form on a single cell or facet of
the mesh. The current production form compiler in FEniCS is the
\emph{FEniCS Form Compiler} (FFC) \cite{Kirby2006,Logg2012ffc} while the
\emph{SyFi Form Compiler} (SFC) \cite{Alnaes2010} was an earlier alternative.
The \emph{UFL Analyser and Compiler System} (UFLACS) has been
developed as a replacement for the previous FFC implementations, and
has recently become default. In this paper we present the \emph{Two-Stage Form Compiler}
(TSFC), a new form compiler in use in the Firedrake project.  The main
motivation in TSFC is to maintain the structure of the input expression as
long as possible to facilitate the application of optimizations at
the highest possible level of abstraction.
A follow-up paper \cite{Homolya2017a} demonstrates this through the
automatic sum factorisation of high-order finite element kernels.
In this paper, however, we focus on explaining how structure can be
maintained through the compilation pipeline.

The contributions of TSFC, and therefore of this paper, are a number of
algorithmic advances over the state of the art form compilers currently
available. By efficiently traversing the input UFL expressions, the compile
time of TSFC scales much better than legacy FFC as form complexity increases and in
the cases where the transformation from the reference cell to the mesh is
non-affine. TSFC preserves the tensor structure of expressions for
longer through the compilation process than either FFC or UFLACS. In particular, this is
achieved by a new approach to isolating the components of a weak form which
must be inserted into different rows and columns of the output tensor.
Finally, TSFC cleanly separates the lowering of finite element constructs
to tensor algebra on the one hand, from the scheduling of those tensor
operations on the other. This facilitates the optimization of this schedule
either in situ or by a downstream tool.

The rest of this work is arranged as follows. In the remainder of this
section we review some concepts necessary for understanding form
compilers, and then give an overview of other current form compilers.
In \cref{sec:structure} we emphasize the importance of preserving
structure, and propose a novel, structure-preserving method for an
issue that arises in discontinuous Galerkin problems.  We compare this
with the approach adopted by UFLACS, and consider their compatibility
with our existing optimization technology.  Then we discuss the first
compilation stage in \cref{sec:ufl2gem}, briefly describing the
preprocessing in UFL,
then explaining the translation of the preprocessed UFL form to the
intermediate tensor algebra language.  \Cref{sec:gem2c} then gives a
detailed description of this intermediate representation along with
its translation to low-level C code.  \Cref{sec:opt} discusses
optimizations employed at various stages of the form compilation
pipeline.  Experimental evaluation is presented in \cref{sec:exp}, and
\cref{sec:conclusion} finally concludes the paper.

\subsection{The role of a form compiler}

Consider the stationary heat equation
$ -\nabla \cdot (\kappa \nabla u) = f $ with $ \kappa $ thermal
conductivity and $ f $ heat source on a domain $ \Omega $ with
$ u = 0 $ on the boundary.  Its standard weak form with test function
$ v \in \hat{V} $ can be written as
\begin{equation}
  \label{eq:weak_form}
  a(v,u) = L(v)
\end{equation}
where $ a $ is a bilinear and $ L $ is a linear form defined as
\begin{align}
  \label{eq:lhs_form}
  a(v,u;x,\kappa) &= \int_\Omega \kappa \nabla u \cdot \nabla v \dx
                    \quad \text{and} \\
  \label{eq:rhs_form}
  L(v;x,f) &= \int_\Omega f v \dx.
\end{align}
Where $ \kappa $ and $ f $ are prescribed spatial functions that
parametrize the forms $ a $ and $ L $, and $ x $ represents the
coordinate field that characterizes the domain $ \Omega $.  Following
the UFL terminology, we refer to $ v $ and $ u $ as \emph{arguments},
and to $ x $, $ \kappa $, and $ f $ as \emph{coefficients}.  The
latter we added in \cref{eq:lhs_form,eq:rhs_form} after a semicolon to
emphasize the presence of form coefficients.  A form must be linear in
its arguments, but not necessarily in its coefficients.

Let $ V $ and $ \hat{V} $ be suitable finite-dimensional
spaces of functions on $ \Omega $ and let their bases be
$ \{ \phi_i \}_{i=1}^s $ and $ \{ \hat{\phi}_i \}_{i=1}^{\hat{s}} $.
Seeking a solution $ u(x) = \sum_{i=1}^s \mathbf{u}_i \phi_i(x) $ in
the solution space $ V $ reduces \cref{eq:weak_form} to a linear
system $ \mathbf{A} \cdot \mathbf{u} = \mathbf{b} $ with
\begin{align}
  \label{eq:lhs_tensor}
  \mathbf{A}_{ij} &= a(\hat{\phi}_i, \phi_j; x, \kappa) \quad \text{and} \\
  \label{eq:rhs_tensor}
  \mathbf{b}_i &= L(\hat{\phi}_i; x, f).
\end{align}
$ \mathbf{A} $ is the bilinear form $ a $ assembled as a matrix, and
$ \mathbf{b} $ is the linear form $ L $ assembled as a vector.

UFL is an embedded domain-specific language for the description of the
weak form and discretized function spaces of finite element problems.
It is embedded in Python. \Cref{lst:ufl} defines $ a $ and $ L $ in
UFL, lines
\ref{lin:lhs} and \ref{lin:rhs} correspond to \cref{eq:lhs_form} and
\cref{eq:rhs_form} respectively.  Line \ref{lin:mesh} defines an
abstract mesh of triangular cells in a 2-dimensional space with
coordinates attached to each vertex (equivalent to a first-order
Lagrange space).  Setting \texttt{dim=3} would create a 2-dimensional
surface in a 3-dimensional space.  Setting the degree of the
coordinate element higher than one would allow the triangular cells to
be curved.  Users of FEniCS and Firedrake typically create a concrete
mesh object which extends the UFL class with mesh data that define a
tesselation of a concrete physical domain.  Line \ref{lin:fs} sets up
the function space $ V = \hat{V} $ with $ P_2 $ elements on the mesh;
we seek a solution in this space.  \Cref{lst:ufl} specifies that
$ \kappa $ and $ f $ are also from $ V $.

\begin{listing}
  \begin{lstlisting}[language=Python,escapechar=|]
mesh = Mesh(VectorElement("Lagrange", triangle, 1, dim=2)) |\label{lin:mesh}|
V = FunctionSpace(mesh, FiniteElement("Lagrange", triangle, 2)) |\label{lin:fs}|

u = TrialFunction(V)
v = TestFunction(V)

kappa = Coefficient(V)
f = Coefficient(V)

a = kappa*dot(grad(u), grad(v))*dx |\label{lin:lhs}|
L = f*v*dx |\label{lin:rhs}|
  \end{lstlisting}
  \caption{Stationary heat equation in UFL}
  \label{lst:ufl}
\end{listing}

The UFL specification serves as the input of a form compiler.  Form
compilation can be understood as the \emph{partial evaluation} of a
multilinear form given a basis for each of its arguments.  For
example, the bilinear form $ a $ is compiled to the function
$ (x, \kappa) \mapsto \mathbf{A} $, and the linear form $ L $ to the
function $ (x, f) \mapsto \mathbf{b} $.  Form compilers typically
assume that the domain consists of a single cell; this makes the
argument bases independent of actual physical domain up to a space
transformation which is encapsulated in the coordinate coefficient
$ x $.  This does not limit the applicability of form compilation
since, given a tessellation $ \mathcal{T} $ of the domain $ \Omega $,
one can evaluate forms locally on each cell and assemble the
contributions into the global tensors efficiently
\cite[\S2.2]{Kirby2006}.

\subsection{A compiler pass}
\label{sec:pullback}

A compiler may be understood as a series of expression
transformations, called \emph{compiler passes}.  A pass takes an
expression and produces a new expression, implementing, for example, a
simplification or an optimization, or translating the expression to a
different language.  A compiler may use several intermediate languages
between the source and target language to decompose the
translation process into several simpler passes, and especially to
facilitate optimizations at the right level of abstraction.

It is common practice in form compilers to change the integral from
physical to reference space.  This enables, for example, the
precomputation of basis functions at quadrature points.  We use this
transformation as an example for a compiler pass.  Let us consider the
Laplace operator
\begin{equation}
  \label{eq:laplace}
  a(v,u) = \int_K \nabla u \cdot \nabla v \dx.
\end{equation}
Let $ x: \tilde{K} \rightarrow K $ be the mapping of coordinates from
the reference cell to the physical cell, and let the Jacobian of this
mapping be $ J = \nabla x $.  For a $ P_2 $ element
$ u(x(X)) = \tilde{u}(X) $, and likewise for $ v $.  The gradient
transforms as $ \nabla_x = J^{-T} \nabla $, and the integral needs
scaling as $ \mathrm{d}x = \left| J \right| \mathrm{d}X $.  Thus we
have
\begin{equation}
  \label{eq:pullback}
  \int_K \nabla u \cdot \nabla v \dx = \int_{\tilde{K}} J^{-T} \nabla
  \tilde{u} \cdot J^{-T} \nabla \tilde{v} \left| J \right| \dX.
\end{equation}

By locally applying the above substitutions to the arguments,
gradients and the integral, we obtain the expression in
\cref{eq:pullback}.  As one can see, $ J^{-T} $ is
referenced twice, and $ J $ is referenced twice directly, or three times
indirectly.  A later pass may replace $ J $ with the
evaluation of $ \nabla x $.  It is therefore better to think of
expressions as \emph{directed acyclic graphs} (DAGs) rather than
trees.  Graph representations of \cref{eq:laplace} and
\cref{eq:pullback} are provided in supplement \cref{sec:ufl_graph}.

\subsection{Facet integrals in discontinuous Galerkin methods}
\label{sec:dg}

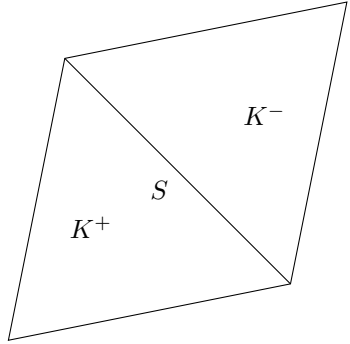
\begin{figure}
  \centering
  \begin{tikzpicture}[scale=0.75]
    \draw (0,0) -- (5,1) -- (6,6) -- (1,5) -- cycle;
    \draw (1,5) -- node[anchor=north east] {$ S $} (5,1);
    \node[left] at (2,2) {$ K^+ $};
    \node[right] at (4,4) {$ K^- $};
  \end{tikzpicture}
  \caption{Two cells $ K^+ $ and $ K^- $ sharing a common facet $ S $.}
  \label{fig:interior_facet}
\end{figure}

Discontinuous Galerkin methods connect cells through flux terms which
are expressed as integrals over the \emph{interior facets} of a mesh.
\O{}lgaard, Logg, and Wells \cite{Oelgaard2008} extended the form
language with interior facet integrals and provided an implementation
in FFC, here we follow their description.  Consider, for example, the
bilinear form
\begin{equation*}
  a(v, u) = \int_S \jump{u} \jump{v} \ds ,
\end{equation*}
where $ \jump{v} $ denotes the \emph{jump} in the function value of
$ v $ across the facet $ S $.  The jump is expressed as
$ \jump{v} = v^+ - v^- $, where $ v^+ $ and $ v^- $ are the function
values as seen from the two cells $ K^+ $ and $ K^- $ incident to
$ S $, respectively (see \cref{fig:interior_facet}).  While UFL
provides utility functions such as \texttt{jump} and \texttt{avg} for
convenience, its only additional primitive is \emph{restriction}:
quantities to be evaluated on interior facets must be restricted to either side of the
facet.  The two sides of a facet are arbitrarily marked as $ + $ and
$ - $.

Let $ \{ \phi^+_i \}_{i=1}^n $ be the local finite element basis on
$ K^+ $, and similarly let $ \{ \phi^-_i \}_{i=1}^n $ be the local
finite element basis on $ K^- $.  Since interior facet integrals
involve both cells, let $ \{ \phi_i \}_{i=1}^{2n} $ be a local basis
on $ K = K^+ \cup K^- $ by the following construction:
\begin{equation}
  \phi_i(x) =
  \begin{cases}
    \phi^+_i(x) & \mbox{if } 1 \le i \le n \mbox{ and } x \in K^+, \\
    0 & \mbox{if } 1 \le i \le n \mbox{ and } x \in K^-, \\
    0 & \mbox{if } n+1 \le i \le 2 n \mbox{ and } x \in K^+, \\
    \phi^-_{i-n}(x) & \mbox{if } n+1 \le i \le 2 n \mbox{ and } x \in K^-.
  \end{cases}
\end{equation}
This construction makes the local tensor
$ \mathbf{A}_{ij} = a(\phi_i, \phi_j) $ a \emph{block tensor}:
\begin{equation}
  \label{eq:block_tensor}
  \mathbf{A} =
  \begin{bmatrix}
    \mathbf{A}^{(+,+)} & \mathbf{A}^{(+,-)} \\
    \mathbf{A}^{(-,+)} & \mathbf{A}^{(-,-)}
  \end{bmatrix}
\end{equation}
FFC, UFLACS and TSFC construct the interior facet tensor
$ \mathbf{A} $ by computing its subtensors.  The subtensor
$ \mathbf{A}^{(+,+)} $ can be seen as the evaluation of a
corresponding form $ a^{(+,+)} $, and the same applies for the other
subtensors.  \Cref{sec:structure} presents a novel algorithm implemented in
TSFC for the separation of the multilinear form into several forms
corresponding to the subtensors of the local tensor.

\subsection{Other contemporary form compilers}
\label{sec:fc}

\begin{figure}
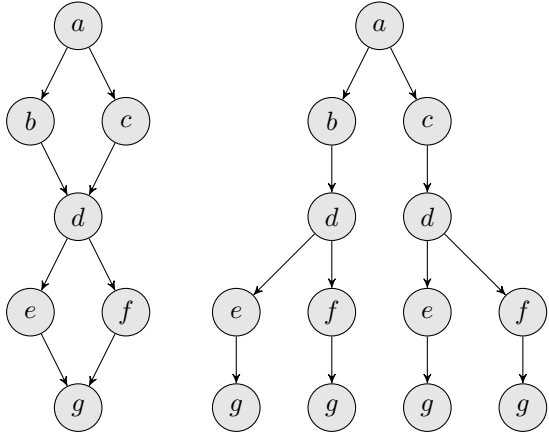

  \centering
  \begin{tikzpicture}[>=stealth',scale=0.5]
    \tikzstyle{n} = [draw,shape=circle,minimum size=1.8em,inner sep=0pt,fill=gray!20]
    \input{ffc_dag}
  \end{tikzpicture}
  \qquad
  \begin{tikzpicture}[>=stealth',scale=0.5]
    \tikzstyle{n} = [draw,shape=circle,minimum size=1.8em,inner sep=0pt,fill=gray!20]
    \input{ffc_dag_explode}
  \end{tikzpicture}
  \caption{DAG representation of an expression with shared
    subexpressions (left), and a \emph{tree view} of the same
    expression (right).}
  \label{fig:ffc}
\end{figure}

FFC contains several compilers for weak forms---they are called
``representations''.  The \emph{quadrature representation}
\cite{Oelgaard2010} generates code that performs runtime quadrature,
akin to what one would manually write.  The \emph{tensor
  representation} \cite{Kirby2006,Kirby2007} computes the local tensor
as the contraction of a precomputable \emph{reference tensor} with an
element-dependent \emph{geometry tensor}.  This allows the elimination
of the quadrature loop, which especially benefits higher-order
schemes.  However, due to the nature of this representation, the size
of the reference tensor grows exponentially with the number of
coefficients in the form, rendering this representation inefficient or
even infeasible for a wide variety of problems.  On the other hand,
the quadrature representation suffers from an implementation
limitation that expressions are seen as a tree rather than a DAG.
This can dramatically increase the perceived size of the input
expression (see \cref{fig:ffc}), which is detrimental to form
compilation speed and generated code size.  In case of non-affine
geometries as well as for ``complicated forms'' --- such as various
formulations of hyperelastic models --- the effect can be
prohibitive.

UFLACS, which was integrated as another FFC representation, and has
recently become the default implementation for most cases, has been
developed to overcome this limitation.  However, to keep the
internal data structures simple, UFLACS is eager to lower high-level
constructs early in the form compilation pipeline.  The high-level
structure expressed in the input form, if preserved longer, could
facilitate optimizations that are otherwise difficult to obtain.
\Cref{sec:structure} provides further insight.

\section{Preserving structure}
\label{sec:structure}

Optimizations are best expressed at a specific level of abstraction.  If the
representation is too high level, the transformation may not be expressible
at all, and if the representation is too low level, difficult analyses may
be needed to recognize the optimization opportunity.  It is therefore
desirable to postpone representation-lowering operations until the
appropriate optimizations have been applied.  There are two principal
ways in which UFLACS lowers structure early, but TSFC does not:
lowering tensor expressions to scalar expressions and \emph{argument
factorization}.

Argument factorization rewrites
the integrand as a sum of products such that each product contains a
term without arguments and for each argument a component of (a
derivative of) that argument.  This is always possible since the form
is multilinear in its arguments.  Lowering to scalars is a practical
requirement: distributivity rules in vector and tensor algebra are
numerous, and completeness of the implementation is necessary for
correctness in all cases.  The sum-of-product form makes it easy to
separate the form into parts that correspond to the subtensors of the
local tensor in interior facet integrals (see \cref{sec:dg}).
We also note that UFLACS applies argument factorization irrespective
of the integral type.  In some cases it turns out to be an effective
optimization as argument-independent expressions are exposed and
precomputed outside the inner loops.  However, this approach is in
general suboptimal.  The transformation may require the expansion of
products involving one or more arguments which affects an expression
in two major ways:
\begin{enumerate*}[label=\textit{\alph*})]
\item by increasing the operation count (which is usually only
  partially counteracted by the subsequent loop hoisting);
  and
\item by destroying the common subexpressions that naturally arise
  when a problem is specified in a tensor algebra language.
\end{enumerate*}

To demonstrate what UFLACS does, let us consider the following
interior facet integral as an example:
\begin{equation}
  \int_S \jump{u}_n \cdot \average{\nabla v} \ds
\end{equation}
where $ \jump{u}_n $ is the product of the jump of $ u $ with the
facet normal $ n $, and $ \average{\nabla v} $ is the average of
$ \nabla v $.  The equivalent integrand with restrictions is
\begin{equation}
  \label{eq:dS_integrand}
  \jump{u}_n \cdot \average{\nabla v} = \frac{1}{2} \left( u^+n^+ +
    u^-n^- \right) \cdot \left({\nabla v}^+ + {\nabla v}^-\right) .
\end{equation}
$ n $ is of unit length, and always pointing out of the cell to which it is
restricted.  UFLACS first lowers the expression to scalars, in this case by
expanding the dot product. Assuming a 2-dimensional space, this step yields
\begin{align*}
  \jump{u}_n \cdot \average{\nabla v}
  &= \frac{1}{2} \sum_{i=1}^2 \left( u^+n_i^+ + u^-n_i^- \right)
    \left(\left({\nabla v^+}\right)_i + \left({\nabla v}^-\right)_i\right) \\
  &= \frac{1}{2} \Big[ \left( u^+n_1^+ + u^-n_1^- \right)
    \left(\left({\nabla v^+}\right)_1 + \left({\nabla v}^-\right)_1\right) \\
  & \qquad
    + \left( u^+n_2^+ + u^-n_2^- \right) \left(\left({\nabla
    v^+}\right)_2 + \left({\nabla v}^-\right)_2\right) \Big].
\end{align*}
Argument factorization is achieved through the application of
distributivity rules:
\begin{multline*}
  \jump{u}_n \cdot \average{\nabla v}
  = \frac{1}{2} u^+n_1^+\left({\nabla v^+}\right)_1 + \frac{1}{2}
    u^+n_1^+\left({\nabla v^-}\right)_1 + \frac{1}{2}
    u^-n_1^-\left({\nabla v^+}\right)_1
  + \frac{1}{2} u^-n_1^-\left({\nabla v^-}\right)_1 \\ +
    \frac{1}{2} u^+n_2^+\left({\nabla v^+}\right)_2 + \frac{1}{2}
    u^+n_2^+\left({\nabla v^-}\right)_2
  + \frac{1}{2} u^-n_2^-\left({\nabla v^+}\right)_2 +
    \frac{1}{2} u^-n_2^-\left({\nabla v^-}\right)_2.
\end{multline*}
At this point one can easily separate the form based on argument
restrictions:
\begin{subequations}
  \label{eq:uflacs}
  \begin{align}
    \label{eq:pp_uflacs}
    (+, +) &\Rightarrow \frac{1}{2} u^+n_1^+\left({\nabla v^+}\right)_1
             + \frac{1}{2} u^+n_2^+\left({\nabla v^+}\right)_2 \\
    \label{eq:pm_uflacs}
    (+, -) &\Rightarrow \frac{1}{2} u^+n_1^+\left({\nabla v^-}\right)_1
             + \frac{1}{2} u^+n_2^+\left({\nabla v^-}\right)_2 \\
    \label{eq:mp_uflacs}
    (-, +) &\Rightarrow \frac{1}{2} u^-n_1^-\left({\nabla v^+}\right)_1
             + \frac{1}{2} u^-n_2^-\left({\nabla v^+}\right)_2 \\
    \label{eq:mm_uflacs}
    (-, -) &\Rightarrow \frac{1}{2} u^-n_1^-\left({\nabla v^-}\right)_1
             + \frac{1}{2} u^-n_2^-\left({\nabla v^-}\right)_2.
  \end{align}
\end{subequations}

In contrast, TSFC separates the form into parts that correspond to the
subtensors without applying argument factorization.  This is achieved
by simply exploiting the linearity of the form in its arguments.
Suppose $ L(v) $ is a linear form and an interior facet integral,
written with restrictions as in \cref{sec:dg}.  Its formulation could
be seen as a function $ \mathcal{F} $ of $ v^+ $ and $ v^- $.
Since each instance of an argument is either positively or negatively
restricted in the form, the following equality holds:
\begin{equation}
  \label{eq:argument_split}
  \mathcal{F}(v^+, v^-) = \mathcal{F}(v^+, 0) + \mathcal{F}(0, v^-) .
\end{equation}
This means that splitting the form based on its
contributions to the subtensors of the interior facet tensor is as
simple as substituting an appropriately shaped zero for any argument
that has the ``wrong'' restriction.  This approach also naturally
generalizes to any number of arguments.  Applying this approach to
\cref{eq:dS_integrand} directly yields:
\begin{subequations}
  \label{eq:tsfc}
  \begin{align}
    \label{eq:pp_tsfc}
    (+, +) &\Rightarrow \frac{1}{2} \left( u^+n^+ + 0 n^- \right) \cdot
             \left({\nabla v}^+ + {\nabla 0}\right) = \frac{1}{2} u^+n^+ \cdot
             {\nabla v}^+ \\
    \label{eq:pm_tsfc}
    (+, -) &\Rightarrow \frac{1}{2} \left( u^+n^+ + 0 n^- \right) \cdot
             \left({\nabla 0} + {\nabla v}^-\right) = \frac{1}{2} u^+n^+ \cdot
             {\nabla v}^- \\
    \label{eq:mp_tsfc}
    (-, +) &\Rightarrow \frac{1}{2} \left( 0 n^+ + u^-n^- \right) \cdot
             \left({\nabla v}^+ + {\nabla 0}\right) = \frac{1}{2} u^-n^- \cdot
             {\nabla v}^+ \\
    \label{eq:mm_tsfc}
    (-, -) &\Rightarrow \frac{1}{2} \left( 0 n^+ + u^-n^- \right) \cdot
             \left({\nabla 0} + {\nabla v}^-\right) = \frac{1}{2} u^-n^- \cdot
             {\nabla v}^-.
  \end{align}
\end{subequations}
Equations \cref{eq:uflacs} and \cref{eq:tsfc} are equivalent, but
TSFC's approach preserves the tensor structure of the original
expression, and does not require rearrangement.  In general, the
preserved structure can be exploited.  For example, COFFEE
\cite{Luporini2015,Luporini2016} explores a \emph{transformation
  space} with the goal of minimising the operation count of a kernel,
and argument factorisation is difficult to undo. This renders
inaccessible to COFFEE a part of the transformation space, potentially
eliminating the most efficient configurations.

Although we cannot presently provide hard evidence for performance
improvement due to preserving the (geometric) tensor structure of the
input form, TSFC is designed to have the option of lowering tensor
structure sooner or later.  Truter \cite{Truter2017} implements
optimizations that must precede the lowering of tensor structure, but
they can be conveniently done before most other form compiler phases.
Homolya, Kirby, and Ham \cite[\S 3.4]{Homolya2017a}, however, show an
example where early lowering to scalars leads to poorer description of
nonzero patterns, but translating this to measurable performance
increase is currently blocked by lack of support for said patterns in
global assembly.

\begin{figure}
  \centering
  \begin{tikzpicture}[>=stealth',scale=1.5]
    \node (ufl) at (0, 0) {UFL};
    \node (c) at (0, -4) {C / COFFEE};

    \node (ufl_) at (-2.5, 0) [gray!100] {\small \textit{Form language}};
    \node (c_) at (-2.5, -4) [gray!100] {\small \textit{Host language}};

    \node (gem) at (0, -2) {GEM};
    \node (gem_) at (-2.5, -2) [gray!100,text width=1in,align=center]
    {\small \textit{Tensor algebra language}};

    \draw[->] (ufl) -> (gem) node [midway,right,text width=1in,align=center]
    {\small lower finite elements and geometry};
    \draw[->] (gem) -> (c) node [midway,right,text width=1in,align=center]
    {\small schedule operations, create loop structure, etc.};
  \end{tikzpicture}
  \caption{TSFC compiles forms in two stages. The first stage
    translates the evaluation of finite element objects to tensor
    algebra expressions. The second stage generates C code to evaluate
    tensor algebra expressions.}
  \label{fig:tsfc_arch}
\end{figure}
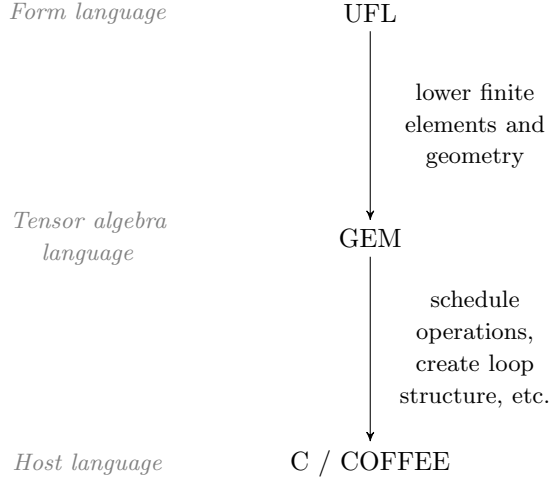

Compared to previous form compilers, we essentially applied an
inversion of order: lowering of finite element objects and geometric
terms happens before either lowering tensor structure or applying
argument factorisation.  The latter are both optional since argument
factorisation does not always improve performance, and explicit loops
for tensor operations might be better than unrolled loops.  This
creates an intermediate representation that is a tensor algebra
expression without any finite element objects or geometric terms, as
sketched in \cref{fig:tsfc_arch}, splitting compilation into two
stages, hence the name: \emph{Two-Stage Form Compiler} (TSFC).
A tensor algebra language (GEM) is used as an intermediate
representation, and the first stage lowers finite element objects and
geometric terms to this language (\cref{sec:ufl2gem}), while the
second stage generates code to evaluate this tensor algebra expression
(\cref{sec:gem2c}).

\section{First stage (UFL $ \boldsymbol{\rightarrow} $ GEM)}
\label{sec:ufl2gem}

GEM is a tensor algebra language very similar to UFL in its tensor algebra
features, although it contains nothing specific to the finite element
method.  This approach works very well, because the evaluation of finite
element objects and geometric terms is expressible using only tensor
algebra.

\subsection{Tensor algebra and index notation in UFL}
\label{sec:free_indices}

The tensor nature of UFL expressions is represented as \emph{shape}
and \emph{free indices}:
\begin{description}
\item[shape] An ordered list of dimensions and their respective
  extent, e.g.\@ (2, 2). A dimension is only identified by its
  position in the shape.
\item[free indices] An unordered set of dimensions where each
  dimension is identified by a symbolic index object.  One might think
  of free indices as an ``unrolled shape''.
\end{description}
These traits are an integral part of any UFL expression.  For example,
let $ B $ be a $ 2 \times 2 $ matrix, then $ B $ has shape (2, 2) and
no free indices. $ B_{1,1} $ has scalar shape and no free indices;
$ B_{i,j} $ has scalar shape and free indices $ i $ and $ j $; and
$ B_{i,1} $ has scalar shape and free index $ i $.
UFL has four basic node types related to the tensor nature of its
expressions:
\begin{itemize}
\item $ \mathtt{Indexed}(A, (\alpha_1, \alpha_2, \ldots, \alpha_r)) $ \\
  where $ A $ is a tensor expression of rank $ r $, and $ \alpha $ is
  multi-index consisting of fixed indices and free indices. The result
  has scalar shape, but all the free indices of $ \alpha $ become free
  indices of the result, i.e.\@ \texttt{Indexed} can convert shape to
  free indices.
\item $ \mathtt{IndexSum}(e, i) $ \\
  Tensor contraction of the scalar-valued expression $e$ over the free index $i$.
\item $ \mathtt{ComponentTensor}(e, \alpha) $ \\
  where $ \alpha := (\alpha_1, \alpha_2, \ldots, \alpha_k) $ is a multi-index consisting of free indices, and
  $ e $ is a scalar-valued expression with free indices
  $ \alpha_1, \alpha_2, \ldots, \alpha_k $ (at least). These free
  indices are turned into shape, in the order as they appear in
  $ \alpha $.
\item $ \mathtt{ListTensor}(e_1, e_2, \ldots, e_n) $ \\
  where $ e_1, e_2, \ldots, e_n $ are expressions of shape
  $ (s_1, s_2, \ldots, s_r ) $. The result is a concatenation of (the
  otherwise unrelated) expressions $ e_1, e_2, \ldots, e_n $, with
  shape $ (n, s_1, s_2, \ldots, s_r ) $.
\end{itemize}

The index notation provides a low-level representation for many vector
and matrix operations.  For example, dot product, element-wise
multiplication and outer product are all expressible using only
multiplication of scalars and some of the above tensor operations:
\begin{itemize}
\item \texttt{dot(u, v)} $ \Rightarrow $ \texttt{IndexSum(u[i] * v[i], i)}
\item \texttt{elem\char`_mult(u, v)} $ \Rightarrow $ \texttt{ComponentTensor(u[i] * v[i], (i,))}
\item \texttt{outer(u, v)} $ \Rightarrow $ \texttt{ComponentTensor(u[i] * v[j], (i, j))}
\end{itemize}

\subsection{UFL preprocessing and modified terminals}
\label{sec:preprocess}

TSFC relies on UFL to preprocess the form.  The most important step in
this procedure is to change the integral from physical to reference
space.  This involves scaling the integral and the application of
pullbacks on the functions (arguments and coefficients) of the form.
For example, the Laplace operator
\begin{equation}
  \label{eq:v_dx}
  \int \nabla u \cdot \nabla v \dx,
\end{equation}
where $ u $ and $ v $ are the trial and the test function
respectively, becomes
\begin{equation}
  \label{eq:v_dX_pullback}
  \int J^{-T} \nabla \tilde{u} \cdot J^{-T} \nabla \tilde{v} \left| J \right| \dX
\end{equation}
if $ v $ and $ u $ require no reference value transformation, as
discussed in \cref{sec:pullback}.  $ J $ is the Jacobian of the
coordinate transformation.  Note the change in the integral measure
($ \mathrm{d}x $ versus $ \mathrm{d}X $).  After this transformation,
UFL does various simplifications on the form: the Jacobian is replaced
with the gradient of the spatial coordinates, determinants are
expanded, divergence and curl are expressed with tensor algebra on
gradients, and various products are expanded using the index notation.
\Cref{sec:free_indices} demonstrated that lowering to the index
notation introduces \texttt{ComponentTensor} nodes; however, to make
this discussion easier to follow, we simplify most of them away, even
though in reality they are only eliminated at a later stage
(\cref{sec:component_tensor}).  Finally, gradients and restrictions
are propagated to \emph{leaf nodes} or \emph{terminals}, i.e.\@ nodes
of the expression graph with no children.  This reduces the number of
language constructs that the form compiler needs to handle while
preserving a rich language for the user.  As a result of these
operations, \cref{eq:v_dX_pullback} becomes:
\begin{align}
  \begin{aligned}
  d &:= \left( \frac{\partial x}{\partial X} \right)_{1,1}
    \left( \frac{\partial x}{\partial X} \right)_{2,2} - \left(
      \frac{\partial x}{\partial X} \right)_{1,2} \left(
    \frac{\partial x}{\partial X} \right)_{2,1} & \triangleright \, \det(J) \\
  B &:= \begin{bmatrix}
    \left( \frac{\partial x}{\partial X} \right)_{2,2} & -\left(
      \frac{\partial x}{\partial X} \right)_{1,2} \\
    -\left( \frac{\partial x}{\partial X} \right)_{2,1} & \left(
      \frac{\partial x}{\partial X} \right)_{1,1}
  \end{bmatrix} & \triangleright \, \mathrm{adj}(J) \\
  K &:= {\bigg] \frac{B_{i_1,i_2}}{d} \bigg[}_{i_1,i_2} &
  \triangleright \, \mathrm{inv}(J) \\
  \int & \sum_{i_5} \left( \sum_{i_3} K_{i_3,i_5} (\nabla
    \tilde{u})_{i_3} \right) \left( \sum_{i_4} K_{i_4,i_5} (\nabla
    \tilde{v})_{i_4} \right) \left| d \right| \dX.
  \end{aligned}
  \label{eq:v_dX}
\end{align}
where $ {]e[}_\alpha $ is a notation for
$ \mathtt{ComponentTensor}(e, \alpha) $.  \Cref{eq:v_dX} denotes a
single expression, although some subexpressions are assigned names for
simpler presentation.

In UFL, terminals are:
\begin{enumerate*}[label=\textit{\alph*})]
\item constant literals,
\item coefficients,
\item arguments,
\item geometric quantities, and
\item quadrature weight.
\end{enumerate*}
Various \emph{terminal modifiers} may wrap a terminal.  TSFC only
needs to handle the following terminal modifiers:
\begin{itemize}
\item \texttt{PositiveRestricted}, \texttt{NegativeRestricted} \\
  Restriction to one particular side of the facet in an interior facet
  integral.
\item \texttt{ReferenceGrad} \\
  Gradient in the reference space, possibly applied multiple times.
\end{itemize}
A \emph{modified terminal} is a terminal with a set of modifiers
attached.
In this set-up, the translation of modified terminals is independent
of their context (surrounding outer expression), and the translation
of the overall expression does not need to know more about the
modified terminals than their shape and free indices.  Since GEM
intentionally mimics the tensor algebra features of UFL, the latter
problem is trivial.  We discuss the former in the next subsection.

\subsection{Substitution of modified terminals}
\label{sec:fem}

Currently, the substitution rules for modified terminals mimic the
evaluation rules of FFC:
\begin{itemize}
\item \textbf{Arguments.} FIAT \cite{Kirby2004} provides the tabulation matrix that
  contains the value of each basis function at each quadrature point
  for the finite element of the argument.  This becomes a compile-time
  constant matrix in GEM.  This matrix is then indexed with the
  quadrature index and the corresponding argument index.
\item \textbf{Coefficients.} We need a GEM expression that evaluates
  the coefficient at each quadrature point.  The matrix-vector product
  of the element tabulation matrix with the local basis function
  coefficients gives exactly that.  That vector for the cell is given
  as a kernel argument, which is represented as a run-time value in
  GEM.
\item \textbf{Geometric quantities.} Because Firedrake represents the mesh
  coordinates as a finite element field, the Jacobian and further
  derivatives of the coordinate field are simply evaluated as coefficients.
  Other geometric terms typically appear in facet integrals (e.g.\@ facet
  normal), and, as a consequence of UFL preprocessing, are expressed in
  reference space. They depend on the cell type, which is known at compile
  time, and often on the facet, which is provided as a kernel argument at
  run time.  Thus the evaluation of a geometric term usually involves
  indexing into a compile-time constant tensor with the facet index.
\end{itemize}

Applying these evaluation rules, the modified terminals of
\cref{eq:v_dX} are replaced with the following expressions:
\begin{align}
  \nabla \tilde{v} &\rightarrow \begin{bmatrix}
    E^{(1)}_{q,j} & E^{(2)}_{q,j}
  \end{bmatrix} \\
  \nabla \tilde{u} &\rightarrow \begin{bmatrix}
    E^{(1)}_{q,k} & E^{(2)}_{q,k}
  \end{bmatrix} \\
  \frac{\partial x}{\partial X} &\rightarrow
                                  \begin{bmatrix}
                                    \sum_r C_{q,r}^{(1,1)}c_r & \sum_r C_{q,r}^{(1,2)}c_r \\
                                    \sum_r C_{q,r}^{(2,1)}c_r & \sum_r C_{q,r}^{(2,2)}c_r
                                  \end{bmatrix}
\end{align}
where $ E^{(b)} $ is the tabulation matrix of the first derivative of
the finite element of $ v $ in the $b$-th direction, $ C^{(a,b)} $ is
the tabulation matrix of the first derivate in the $b$-th direction of
the $a$-th component of the coordinate element, and $ c $ is the
vector of the local basis function coefficients of the coordinates,
which is not known at compile time but provided as a run-time
parameter.  The quadrature index is $q$, while $j$ and $k$ are the
test and trial function indices respectively, and $r$ is the basis
function index of the coordinate element.

This general treatment of the Jacobian also works for non-simplex
cells and higher-order geometries.  We will discuss optimizations for
affine simplices in \cref{sec:opt}.  Since indexing a tensor with
fixed indices immediately selects the appropriate entry, these
substitutions yield the following integrand expression:
\begin{align*}
  \begin{aligned}
    \label{eq:jacobian}
    J^{(1,1)} &:= \sum_r C_{q,r}^{(1,1)}c_r \\
    J^{(1,2)} &:= \sum_r C_{q,r}^{(1,2)}c_r \\
    J^{(2,1)} &:= \sum_r C_{q,r}^{(2,1)}c_r \\
    J^{(2,2)} &:= \sum_r C_{q,r}^{(2,2)}c_r
  \end{aligned}
  &&
  \begin{aligned}
    d &:= J^{(1,1)} J^{(2,2)} - J^{(1,2)} J^{(2,1)} \\
    B &:= \begin{bmatrix}
      J^{(2,2)} & -J^{(1,2)} \\
      -J^{(2,1)} & J^{(1,1)}
    \end{bmatrix} \\
    K &:= {\bigg] \frac{B_{i_1,i_2}}{d} \bigg[}_{i_1,i_2}
  \end{aligned}
\end{align*}
\begin{equation*}
  \int \sum_{i_5} \left( \sum_{i_3} K_{i_3,i_5} \begin{bmatrix}
      E^{(1)}_{q,k} & E^{(2)}_{q,k}
    \end{bmatrix}_{i_3} \right)
  \left( \sum_{i_4} K_{i_4,i_5} \begin{bmatrix}
      E^{(1)}_{q,j} & E^{(2)}_{q,j}
    \end{bmatrix}_{i_4} \right) \left| d \right| \dX.
\end{equation*}
Aggregating the integrand values for each quadrature point with
quadrature weight $ w_q $ gives the GEM representation of
\cref{eq:v_dx} as
\begin{equation}
  \label{eq:gem}
  \sum_q w_q \left| d \right| \sum_{i_5} \left( \sum_{i_3}
    K_{i_3,i_5} \begin{bmatrix}
      E^{(1)}_{q,k} & E^{(2)}_{q,k}
    \end{bmatrix}_{i_3} \right)
  \left( \sum_{i_4} K_{i_4,i_5} \begin{bmatrix}
      E^{(1)}_{q,j} & E^{(2)}_{q,j}
    \end{bmatrix}_{i_4} \right) .
\end{equation}
This representation only contains tensor algebra operators applied to
compile-time constant tensors and a run-time argument to the local kernel.
All the finite element and geometric terms have been lowered, the first
stage of form compilation is complete.

\subsection{Intermediate representation (GEM)}
\label{sec:ir}

Since GEM mimics the tensor algebra features of UFL and lacks other
features, such as finite element objects and geometric quantities, one may
rightly ask why the tensor algebra subset of UFL could not be adopted as the
intermediate representation.  The reason for the creation of GEM is to relax
some of the constraints on \emph{free indices} which are intrinsic in
UFL.

Free indices are just an ``unrolled shape'' in UFL, that may be
``rolled back'' any time.  For any operation, free indices of the
operands are subject to the same constraints that apply to shape. For
example, tensors must have equal shape for addition, consequently,
they also must have the same free indices.  GEM interprets free
indices differently.  A free index in GEM simply means that the value
of the expression depends on that index.  For example, let $ u $ be a
3-dimensional vector (no free indices). $ u_i $ and $ u_j $ are
scalars with free indices $i$ and $j$, respectively.  So far, it is
identical to UFL. However, in GEM $ u_i + u_j $ is just an expression
that depends on both $i$ and $j$, that is a scalar with free indices
$i$ and $j$, while in UFL the same expression is illegal because the
free indices do not match: the addition of a 3-dimensional vector with
itself cannot produce a $ 3 \times 3 $ matrix.

While the enforcement of strict shape checking is valuable for
preventing user errors in a weak form, this strictness becomes an
obstacle when the modified terminals are replaced with their
respective evaluation rules.  For example, we replaced $ \nabla
\tilde{v} $ with $ \begin{bmatrix}
  E^{(1)}_{q,j} & E^{(2)}_{q,j}
\end{bmatrix} $ in \cref{sec:fem}.  In UFL this would mean replacing a
vector with a third-order tensor, which yields an illegal expression in the
general case.  In GEM it just replaces a vector with another vector
which depends on $ q $ (quadrature index) and $ j $ (argument index).
There is eventually a summation over $ q $, and the dependence on
$ j $ gives the components of the element tensor.  This difference in
free indices poses no difficulty for the UFL $ \rightarrow $ GEM
translation since GEM is strictly more permissive.

\section{Second stage (GEM $ \boldsymbol{\rightarrow} $ C)}
\label{sec:gem2c}

This section discusses the second stage of form
compilation, which is the translation of GEM to C, a programming
language with explicit loops and only scalar expressions.

\subsection{Overview of GEM}
\label{sec:gem}

GEM was intentionally designed to be very similar to UFL in its tensor
algebra features.  In common with UFL, the tensor nature of expressions
can be expressed as \emph{shape} or \emph{free indices} (see
\cref{sec:free_indices}).  Most supported operations are naturally
scalar, such as addition and multiplication, but there are a few
tensor operations as well, such as tensor contraction (called
\texttt{IndexSum}).  Here we give a summary of all GEM nodes:

\begin{itemize}
\item \textbf{Terminals:}
  \begin{itemize}
  \item \texttt{Literal} (tensor literal)
  \item \texttt{Zero} (all-zero tensor)
  \item \texttt{Identity} (identity matrix)
  \item \texttt{Variable} (run-time value, for kernel arguments)
  \end{itemize}
\item \textbf{Scalar operations:}
  \begin{itemize}
  \item \textit{Binary operators:} \texttt{Sum}, \texttt{Product},
    \texttt{Division}, \texttt{Power}, \texttt{MinValue}, \texttt{MaxValue}
  \item \textit{Unary operators:} \texttt{MathFunction} (e.g.\@ $ \sin $, $ \cos $)
  \item \texttt{Comparison} ($>$, $\ge$, $=$, $\ne$, $<$, $\le$):
    compares numbers, returns Boolean
  \item \textit{Logical operators:} \texttt{LogicalAnd},
    \texttt{LogicalOr}, \texttt{LogicalNot}
  \item \texttt{Conditional}: selects between a ``true'' and a
    ``false'' expression based on a Boolean valued \emph{condition}
   \end{itemize}
\item \textbf{Index types:}  
  \begin{itemize}
  \item \texttt{int} (fixed index)
  \item \texttt{Index} (free index): creates a loop at code generation
  \item \texttt{VariableIndex} (unknown fixed index): index value only
    known at run-time, e.g.\@ facet number
 \end{itemize}
\item \textbf{Tensor nodes:} \texttt{Indexed},
  \texttt{ComponentTensor}, \texttt{IndexSum},
  \texttt{ListTensor} \\
  See \cref{sec:free_indices} for details.
\end{itemize}

\subsection{Elimination of \texttt{ComponentTensor} nodes}
\label{sec:component_tensor}

The duplicity of \emph{shape} and \emph{free indices} would noticeably
complicate the process of code generation.  To avoid that, we
eliminate non-scalar shape in non-terminal nodes by removing all
\texttt{ComponentTensor} nodes using a rule that is analogous to
\emph{beta reduction}:
\begin{equation}
  \label{eq:component_tensor}
  \mathtt{Indexed}(\mathtt{ComponentTensor}(e, \alpha), \beta)
  \rightarrow e \vert_{\alpha \rightarrow \beta}
\end{equation}
where $ e $ is an expression with scalar shape, $ \alpha $ is a
multi-index containing distinct free indices of $ e $, and $ \beta $
is a multi-index of the same rank as $ \alpha $.
$ e \vert_{\alpha \rightarrow \beta} $ denotes a substitution of
indices in $ e $, where $ \alpha_i $ is replaced with $ \beta_i $ for
all $ i $.  The rule replaces indexing into a \texttt{ComponentTensor}
by a substitution of the indices.

A \texttt{ComponentTensor} node must be immediately inside an
\texttt{Indexed} node or at the root of the expression because
\texttt{Indexed} is the only GEM operator that accepts non-scalar
shape.  Since
the UFL $ \rightarrow $ GEM stage produces a GEM expression with
scalar shape and the argument indices as free indices, the application
of \cref{eq:component_tensor} removes all \texttt{ComponentTensor}s
from the intermediate representation.  The downside of this approach
is the possibility of code duplication when different substitutions
are applied on the same subexpression.

To demonstrate this, we continue with our Laplace operator example.
To simplify our presentation in \cref{sec:preprocess}, most
\texttt{ComponentTensor} nodes were eliminated except for $ K $
defined in \cref{eq:v_dX}.  Eliminating $ K $ in \cref{eq:gem} gives
\begin{equation}
  \label{eq:gem_noct}
  \sum_q w_q \left| d \right| \sum_{i_5} \left( \sum_{i_3}
    \frac{B_{i_3,i_5}}{d} \begin{bmatrix}
      E^{(1)}_{q,k} & E^{(2)}_{q,k}
    \end{bmatrix}_{i_3} \right)
  \left( \sum_{i_4} \frac{B_{i_4,i_5}}{d} \begin{bmatrix}
      E^{(1)}_{q,j} & E^{(2)}_{q,j}
    \end{bmatrix}_{i_4} \right).
\end{equation}
As one can see, this step duplicates the element-wise division of the
adjugate matrix by the determinant.

\subsection{Code generation}
\label{sec:codegen}

Since \texttt{ComponentTensor} nodes are eliminated, we can now
restrict our attention to the index notation.  A simple code
generation approach for the evaluation of such tensor expressions is
to introduce a \emph{temporary variable} and generate a pair of
\emph{perfect loop nests} (\cref{fig:perfect_loop_nest}) for each
tensor contraction (\texttt{IndexSum}).  The first loop nest
initializes the temporary, and the second accumulates the result.
Consequently, the first loop nest contains all the free indices of the
\texttt{IndexSum} expression, while the second loop nest has one more
index, the index being summed over.  For example, applying this
approach to \cref{eq:gem_noct}, we introduce temporaries
$ J^{(1,1)} $, $ J^{(1,2)} $, $ J^{(2,1)} $, and $ J^{(2,2)} $ as on
page \pageref{eq:jacobian}, $ t_1 $, $ t_2 $, and $ t_3 $ for $
\sum_{i_3} $, $ \sum_{i_4} $, and $ \sum_{i_5} $ respectively, and $ A
$ for the whole expression.
(The code for this intermediate step is provided in the supplement in
\cref{alg:simple}.)

\begin{figure}
  \centering
    \begin{subfigure}[t]{0.45\textwidth}
    \begin{algorithmic}[1]
      \ForAll {$ i $}
      \ForAll {$ j $}
      \State $ d[i, j] \leftarrow (a[i] + b[i]) * c[j] $
      \EndFor
      \EndFor
      \item[]
    \end{algorithmic}
    \caption{A \emph{perfect loop nest} of two loops: the inner loop
      is perfectly nested in the outer loop since the outer loop
      contains no other statements than the inner loop.}
    \label{fig:perfect_loop_nest}
  \end{subfigure}
  \hfill
  \begin{subfigure}[t]{0.45\textwidth}
    \begin{algorithmic}[1]
      \ForAll {$ i $}
      \State $ t[i] \leftarrow a[i] + b[i] $
      \ForAll {$ j $}
      \State $ d[i, j] \leftarrow t[i] * c[j] $
      \EndFor
      \EndFor
    \end{algorithmic}
    \caption{The result of hoisting operations from the inner loop; an
      \emph{imperfect} loop nest.}
    \label{fig:imperfect_loop_nest}
  \end{subfigure}

  \caption{An example of \emph{loop hoisting.}  The calculation
    $ a[i] + b[i] $ is independent of the inner loop, thus it need not
    be repeated for all $ j $.}
  \label{fig:loop_hoisting}
\end{figure}

While this simple approach produces correct code, introducing
temporaries merely for tensor contractions may cause too much
\emph{inlining}.  In particular, shared subexpressions other than
tensor contractions are still expanded, and there is also room for
\emph{loop hoisting} (\cref{fig:loop_hoisting}).  Therefore, we also
introduce temporaries to evaluate subexpressions that
\begin{enumerate*}[label=\textit{\alph*})]
\item \label{enm:refcount} have a \emph{reference count} greater than
  one, or
\item \label{enm:indices} have a different set of \emph{free indices}
  than their unique parent.
\end{enumerate*}
In our example, the determinant of the Jacobian is an instance of
\ref{enm:refcount}, and $ w_q * \mathrm{abs}(\ldots) $ in the final
loop nest is an instance of \ref{enm:indices}. So we also introduce
temporaries $ d $ and $ t_4 $ respectively.  The result of this is
\cref{alg:unfused}, which also adds loop nests for ``defining''
\texttt{ListTensor} expressions, such as adjugate matrix $ B $,
$ l_1 = \begin{bmatrix}
  E^{(1)}_{q,k} & E^{(2)}_{q,k}
\end{bmatrix} $, and $ l_2 = \begin{bmatrix}
  E^{(1)}_{q,j} & E^{(2)}_{q,j}
\end{bmatrix} $.

\begin{algorithm}
  \begin{multicols}{2}
  \begin{algorithmic}[1]
    \ForAll {$ q $}
    \State $ J^{(1,1)}[q] \leftarrow 0 $
    \EndFor
    \ForAll {$ q, r $}
    \State $ J^{(1,1)}[q] $ += $ C_{q,r}^{(1,1)} * c_r $
    \EndFor
    \ForAll {$ q $}
    \State $ J^{(1,2)}[q] \leftarrow 0 $
    \EndFor
    \ForAll {$ q, r $}
    \State $ J^{(1,2)}[q] $ += $ C_{q,r}^{(1,2)} * c_r $
    \EndFor
    \ForAll {$ q $}
    \State $ J^{(2,1)}[q] \leftarrow 0 $
    \EndFor
    \ForAll {$ q, r $}
    \State $ J^{(2,1)}[q] $ += $ C_{q,r}^{(2,1)} * c_r $
    \EndFor
    \ForAll {$ q $}
    \State $ J^{(2,2)}[q] \leftarrow 0 $
    \EndFor
    \ForAll {$ q, r $}
    \State $ J^{(2,2)}[q] $ += $ C_{q,r}^{(2,2)} * c_r $
    \EndFor
    \ForAll {$ q $}
    \State $ B[q, \ldots] \leftarrow {\small \begin{bmatrix}
      J^{(2,2)}[q] & -J^{(1,2)}[q] \\
      -J^{(2,1)}[q] & J^{(1,1)}[q]
    \end{bmatrix}} $
    \EndFor
    \ForAll {$ q, k $}
    \State $ l_1[q, k, \ldots] \leftarrow \begin{bmatrix}
      E^{(1)}_{q,k} & E^{(2)}_{q,k}
    \end{bmatrix} $
    \EndFor
    \ForAll {$ i_5, q, k $}
    \State $ t_1[i_5, q, k] \leftarrow 0 $
    \EndFor
    \ForAll {$ q $}
    \Comment{{\small Common subexpression}}
    \State $ d[q] = J^{(1,1)}[q] * J^{(2,2)}[q] - J^{(1,2)}[q] * J^{(2,1)}[q] $
    \EndFor
    \ForAll {$ i_3, i_5, q, k $}
    \State $ t_1[i_5, q, k] $ += $ B[q, i_3, i_5] \div d[q] * l_1[q, k, i_3] $
    \EndFor
    \ForAll {$ q, j $}
    \State $ l_2[q, j, \ldots] \leftarrow \begin{bmatrix}
      E^{(1)}_{q,j} & E^{(2)}_{q,j}
    \end{bmatrix} $
    \EndFor
    \ForAll {$ i_5, q, j $}
    \State $ t_2[i_5, q, j] \leftarrow 0 $
    \EndFor
    \ForAll {$ i_4, i_5, q, j $}
    \State $ t_2[i_5, q, j] $ += $ B[q, i_4, i_5] \div d[q] * l_2[q, j, i_4] $
    \EndFor
    \ForAll {$ q, j, k $}
    \State $ t_3[q, j, k] \leftarrow 0 $
    \EndFor
    \ForAll {$ i_5, q, j, k $}
    \State $ t_3[q, j, k] $ += $ t_1[i_5, q, k] * t_2[i_5, q, j] $
    \EndFor
    \ForAll {$ j, k $}
    \State $ A[j, k] \leftarrow 0 $
    \EndFor
    \ForAll {$ q $}
    \Comment{Hoisted computation}
    \State $ t_4[q] = w_q * \mathrm{abs}(d[q]) $
    \EndFor
    \ForAll {$ j, k, q $}
    \State $ A[j, k] $ += $ t_4[q] * t_3[q, j, k] $
    \EndFor
  \end{algorithmic}
  \end{multicols}
  \caption{Generated loop nests for the expression \cref{eq:gem_noct}.}
  \label{alg:unfused}
\end{algorithm}

Finally, we apply \emph{loop fusion} to reduce the size of the
temporaries (see \cref{fig:loop_fusion}).  Here we use a simple and
fast algorithm.  First, we construct an ordering of all free indices
that appear in the GEM expression, and apply this ordering to each
(perfect) loop nest.  Based on our knowledge about the structure of
finite element assembly kernels, and in common with FFC, we choose the
\emph{quadrature index} to be the first (outermost), followed by the
\emph{argument indices} in order, and then by all other indices.  A
possible ordering that we use for our example is
$ (q, j, k, r, i_3, i_4, i_5) $.  This means a number of \emph{loop
  interchanges} (\cref{fig:loop_interchange}) in \cref{alg:unfused}.
Considering dependencies, we also take a \emph{topological ordering}
of the loop nests that heuristically tries to maximize loop fusion
(\cref{sec:scheduling}).  For our example, this results in
\cref{alg:reordered}.  Finally, we apply maximal loop fusion without
further reordering.  The fused algorithm is \cref{alg:fused}.  Note
that many of the temporaries are now scalar, or at least have a
reduced tensor order.  The supplement contains a transcription to C
code in \cref{lst:kernel}.

\begin{figure}
  \centering
  \begin{subfigure}[t]{0.3\textwidth}
    \begin{algorithmic}[1]
      \ForAll {$ i $}
      \State $ c[i] \leftarrow a[i] + b[i] $
      \EndFor
      \ForAll {$ i $}
      \State $ d[i] \leftarrow c[i] * c[i] $
      \EndFor
    \end{algorithmic}
    \caption{Sequence of two loops}
  \end{subfigure}
  \hfill
  \begin{subfigure}[t]{0.3\textwidth}
    \begin{algorithmic}[1]
      \ForAll {$ i $}
      \State $ c[i] \leftarrow a[i] + b[i] $
      \State $ d[i] \leftarrow c[i] * c[i] $
      \EndFor
      \item[]
    \end{algorithmic}
    \caption{After loop fusion}
  \end{subfigure}
  \hfill
  \begin{subfigure}[t]{0.3\textwidth}
    \begin{algorithmic}[1]
      \ForAll {$ i $}
      \State $ c \leftarrow a[i] + b[i] $
      \State $ d[i] \leftarrow c * c $
      \EndFor
      \item[]
    \end{algorithmic}
    \caption{After shape reduction}
  \end{subfigure}

  \caption{An example of a pair of loops before and after \emph{loop
      fusion.}  If $ c $ is not needed later, its shape can be reduced
    after fusion.}
  \label{fig:loop_fusion}
\end{figure}

\begin{figure}
  \centering
    \begin{subfigure}[t]{0.45\textwidth}
    \begin{algorithmic}[1]
      \ForAll {$ i $}
      \ForAll {$ j $}
      \State $ c[j, i] \leftarrow a[i] * b[j] $
      \EndFor
      \EndFor
    \end{algorithmic}
  \end{subfigure}
  \hfill
  \begin{subfigure}[t]{0.45\textwidth}
    \begin{algorithmic}[1]
      \ForAll {$ j $}
      \ForAll {$ i $}
      \State $ c[j, i] \leftarrow a[i] * b[j] $
      \EndFor
      \EndFor
    \end{algorithmic}
  \end{subfigure}

  \caption{An example of \emph{loop interchange}: the inner and outer
    loops are swapped.  Loop interchange is only applicable to a perfect loop nest (see
    \cref{fig:perfect_loop_nest}).}
  \label{fig:loop_interchange}
\end{figure}

\begin{algorithm}
  \begin{multicols}{2}  
  \begin{algorithmic}[1]
    \ForAll {$ j, k $}
    \State $ A[j, k] \leftarrow 0 $
    \EndFor
    \ForAll {$ q $}
    \State $ J^{(1,1)}[q] \leftarrow 0 $
    \EndFor
    \ForAll {$ q $}
    \State $ J^{(1,2)}[q] \leftarrow 0 $
    \EndFor
    \ForAll {$ q $}
    \State $ J^{(2,1)}[q] \leftarrow 0 $
    \EndFor
    \ForAll {$ q $}
    \State $ J^{(2,2)}[q] \leftarrow 0 $
    \EndFor
    \ForAll {$ q, r $}
    \State $ J^{(1,1)}[q] $ += $ C_{q,r}^{(1,1)} * c_r $
    \EndFor
    \ForAll {$ q, r $}
    \State $ J^{(1,2)}[q] $ += $ C_{q,r}^{(1,2)} * c_r $
    \EndFor
    \ForAll {$ q, r $}
    \State $ J^{(2,1)}[q] $ += $ C_{q,r}^{(2,1)} * c_r $
    \EndFor
    \ForAll {$ q, r $}
    \State $ J^{(2,2)}[q] $ += $ C_{q,r}^{(2,2)} * c_r $
    \EndFor
    \ForAll {$ q $}
    \State $ B[q, \ldots] \leftarrow {\small \begin{bmatrix}
      J^{(2,2)}[q] & -J^{(1,2)}[q] \\
      -J^{(2,1)}[q] & J^{(1,1)}[q]
    \end{bmatrix}} $
    \EndFor
    \ForAll {$ q $}
    \State $ d[q] = J^{(1,1)}[q] * J^{(2,2)}[q] - J^{(1,2)}[q] * J^{(2,1)}[q] $
    \EndFor
    \columnbreak
    \ForAll {$ q, k $}
    \State $ l_1[q, k, \ldots] \leftarrow \begin{bmatrix}
      E^{(1)}_{q,k} & E^{(2)}_{q,k}
    \end{bmatrix} $
    \EndFor
    \ForAll {$ q, k, i_5 $}
    \State $ t_1[q, k, i_5] \leftarrow 0 $
    \EndFor
    \ForAll {$  q, k, i_3, i_5 $}
    \State $ t_1[q, k, i_5] $ += $ B[q, i_3, i_5] \div d[q] * l_1[q, k, i_3] $
    \EndFor
    \ForAll {$ q $}
    \State $ t_4[q] = w_q * \mathrm{abs}(d[q]) $
    \EndFor
    \ForAll {$ q, j $}
    \State $ l_2[q, j, \ldots] \leftarrow \begin{bmatrix}
      E^{(1)}_{q,j} & E^{(2)}_{q,j}
    \end{bmatrix} $
    \EndFor
    \ForAll {$ q, j, i_5 $}
    \State $ t_2[q, j, i_5] \leftarrow 0 $
    \EndFor
    \ForAll {$ q, j, i_4, i_5 $}
    \State $ t_2[q, j, i_5] $ += $ B[q, i_4, i_5] \div d[q] * l_2[q, j, i_4] $
    \EndFor
    \ForAll {$ q, j, k $}
    \State $ t_3[q, j, k] \leftarrow 0 $
    \EndFor
    \ForAll {$ q, j, k, i_5 $}
    \State $ t_3[q, j, k] $ += $ t_1[q, k, i_5] * t_2[q, j, i_5] $
    \EndFor
    \ForAll {$ q, j, k $}
    \State $ A[j, k] $ += $ t_4[q] * t_3[q, j, k] $
    \EndFor
    \vfill\null
  \end{algorithmic}
  \end{multicols}
  \caption{Reordering loops in \cref{alg:unfused}.}
  \label{alg:reordered}
\end{algorithm}

\begin{algorithm}
  \begin{algorithmic}[1]
    \ForAll {$ j, k $}
    \State $ A[j, k] \leftarrow 0 $
    \EndFor
    \ForAll {$ q $}
    \State $ J^{(1,1)} \leftarrow 0 $
    \State $ J^{(1,2)} \leftarrow 0 $
    \State $ J^{(2,1)} \leftarrow 0 $
    \State $ J^{(2,2)} \leftarrow 0 $
    \ForAll {$ r $}
    \State $ J^{(1,1)} $ += $ C_{q,r}^{(1,1)} * c_r $
    \State $ J^{(1,2)} $ += $ C_{q,r}^{(1,2)} * c_r $
    \State $ J^{(2,1)} $ += $ C_{q,r}^{(2,1)} * c_r $
    \State $ J^{(2,2)} $ += $ C_{q,r}^{(2,2)} * c_r $
    \EndFor
    \State $ B = \begin{bmatrix}
      J^{(2,2)} & -J^{(1,2)} \\
      -J^{(2,1)} & J^{(1,1)}
    \end{bmatrix} $
    \State $ d = J^{(1,1)} * J^{(2,2)} - J^{(1,2)} * J^{(2,1)} $
    \ForAll {$ k $}
    \State $ l_1 \leftarrow \begin{bmatrix}
      E^{(1)}_{q,k} & E^{(2)}_{q,k}
    \end{bmatrix} $
    \ForAll {$ i_5 $}
    \State $ t_1[k, i_5] \leftarrow 0 $
    \EndFor
    \ForAll {$ i_3, i_5 $}
    \State $ t_1[k, i_5] $ += $ B[i_3, i_5] \div d * l_1[i_3] $
    \EndFor
    \EndFor
    \State $ t_4 = w_q * \mathrm{abs}(d) $
    \ForAll {$ j $}
    \State $ l_2 \leftarrow \begin{bmatrix}
      E^{(1)}_{q,j} & E^{(2)}_{q,j}
    \end{bmatrix} $
    \ForAll {$ i_5 $}
    \State $ t_2[i_5] \leftarrow 0 $
    \EndFor
    \ForAll {$ i_4, i_5 $}
    \State $ t_2[i_5] $ += $ B[i_4, i_5] \div d * l_2[i_4] $
    \EndFor
    \ForAll {$ k $}
    \State $ t_3 \leftarrow 0 $
    \ForAll {$ i_5 $}
    \State $ t_3 $ += $ t_1[k, i_5] * t_2[i_5] $
    \EndFor
    \State $ A[j, k] $ += $ t_4 * t_3 $
    \EndFor
    \EndFor
    \EndFor
  \end{algorithmic}
  \caption{The result of loop fusion on \cref{alg:reordered}.}
  \label{alg:fused}
\end{algorithm}

\subsection{Topological orderings of loop nests}
\label{sec:scheduling}

The GEM DAG describes the dependencies of each node in that DAG. The nodes
can be evaluated in any order which respects those dependencies. That is, if
there is an edge from node $a$ to node $b$ then node $b$ must be evaluated
before node $a$. Any ordering which respects the constraints imposed by the
DAG edges is called a \emph{topological ordering}, and the process of
generating such an ordering is called \emph{topological sorting}.

\emph{Topological sorting} is a well-understood problem \cite[\S
22.4]{Cormen2001}. Every DAG admits a topological ordering, but this is not
in general unique.  For correct code generation as described in
\cref{sec:codegen}, any topological ordering suffices.  However, the choice
of ordering determines which loops can be fused, and thus how much memory is
saved.  Reducing the \emph{working set size}\footnote{The amount of memory
  required to carry out a sequence of operations.}  can help to fit into
faster memory, thus also reducing execution time.

We use a variant of Kahn's algorithm \cite{Kahn1962} \cite[\S 2.2.3,
  Algorithm T]{knuth1968art} with
heuristics to maximize loop fusion. \Cref{alg:kahn} presents Kahn's algorithm for
topological sorting.  Line \ref{toposort:choose-element} is a point of ambiguity, since $ S $ can
have more than one element.  Any choice will yield a correct, but
different, topological ordering, so our heuristics are about choosing
``wisely''.

\begin{algorithm}
  \begin{algorithmic}[1]
    \Function {Topological-Sort}{$ G $}
      \State $ \mathit{refcount} \leftarrow $ in-degree for each node in $ G $
      \State $ L \leftarrow [ \, ] $
      \State $ S \leftarrow \{ u \mid \mathit{refcount}[u] = 0 \} $
      \Comment root nodes of $ G $
      \While {$ S $ is non-empty}
        \State $ u \leftarrow \textsc{Choose-Element}(S) $ \label{toposort:choose-element}
        \State $ S \leftarrow S - \{ u \} $
        \State insert $ u $ to the front of $ L $
        \ForAll {$ v \in \mathit{children}[u] $}
          \State $ \textsc{Decrement}(\mathit{refcount}[v]) $
          \If {$ \mathit{refcount}[v] = 0 $}
            \State $ S \leftarrow S \cup \{ v \} $
          \EndIf
        \EndFor
      \EndWhile
      \State \Return $ L $
    \EndFunction
  \end{algorithmic}
  \caption{Kahn's algorithm for \textit{topological sorting}}
  \label{alg:kahn}
\end{algorithm}

Loop fusion occurs between two consecutive loop nests for all common
outer loops.  We maximize this at every choice in a greedy manner.
The number of loops in the previously chosen loop nest, however, poses
an upper limit on the number of loops that can be fused as a result of
the current choice.  For example, if the previous choice was a loop
nest with indices $ (q,) $, this rule so far expresses no preference between
$ (q,) $ and $ (q,r) $ for the next choice, because only the $ q $
loop can be fused in both cases.  Therefore we further refine our
heuristic to prefer those loop nests with exactly the same loops as
the previously chosen loop nest.  Delaying ``entry'' into further
inner loops, is a heuristic designed in the hope that more loop nests with those indices will
satisfy the dependency requirements once we have no other option but
to \emph{do} ``enter'' those further inner loops, thus enabling more
loop fusion later.

\begin{algorithm}
  \begin{algorithmic}[1]
    \Function {Choose-Element}{$ S $}
      \While {\textbf{true}}
      \State $ S_1 = \{ u \in S \mid \mathit{loop\_indices}[u] =
      \mathit{cursor} \} $  \label{choose:pick-start}
      \If {$ S_1 $ is non-empty}
        \State \Return any element from $ S_1 $ \label{choose:pick-end}
      \EndIf
      \State $ S_2 = \{ u \in S \mid \mathit{cursor} \mbox{ is a
        prefix of } \mathit{loop\_indices}[u] \} $ \label{choose:prefix-start}
      \If {$ S_2 $ is non-empty}
        \State $ u \leftarrow $ any element from $ S_2 $
        \State $ \mathit{cursor} \leftarrow \mathit{loop\_indices}[u] $
        \State \Return $ u $ \label{choose:prefix-end}
      \EndIf
      \State remove last index from $ \mathit{cursor} $
      \EndWhile
    \EndFunction
  \end{algorithmic}
  \caption{Heuristic to maximize loop fusion}
  \label{alg:heuristic}
\end{algorithm}

\Cref{alg:heuristic} shows our version of the
\textsc{Choose-Element} function.  We maintain the ordered loop
indices of the previously chosen loop nest in $ \mathit{cursor} $,
which is initialized to $ () $ at the beginning of a topological sort.
We can assume that $ S $ is non-empty.  First, we try to choose an
element with the same loop indices as the previous one (lines \ref{choose:pick-start}--\ref{choose:pick-end}).
If that is not possible, we try to choose an element such that
$ \mathit{cursor} $ is a prefix of its loop indices (lines \ref{choose:prefix-start}--\ref{choose:prefix-end}), and
we update $ \mathit{cursor} $ accordingly.  If that is not possible
either, we drop the last (innermost) index from $ \mathit{cursor} $
and repeat.  Before this step $ \mathit{cursor} $ must have had at least one
index, otherwise $ S_2 = S \ne \varnothing $, so the function would
have returned in line \ref{choose:prefix-end}.

\begin{figure}
  \centering
    \begin{tikzpicture}[>=stealth',scale=0.75,every node/.style={scale=0.80}]
    \tikzstyle{n} = [draw,shape=ellipse,minimum size=2.4em,inner sep=0pt,fill=gray!20]
    \tikzstyle{r} = [draw,minimum size=2.4em,inner sep=0pt,fill=gray!50]
    \input{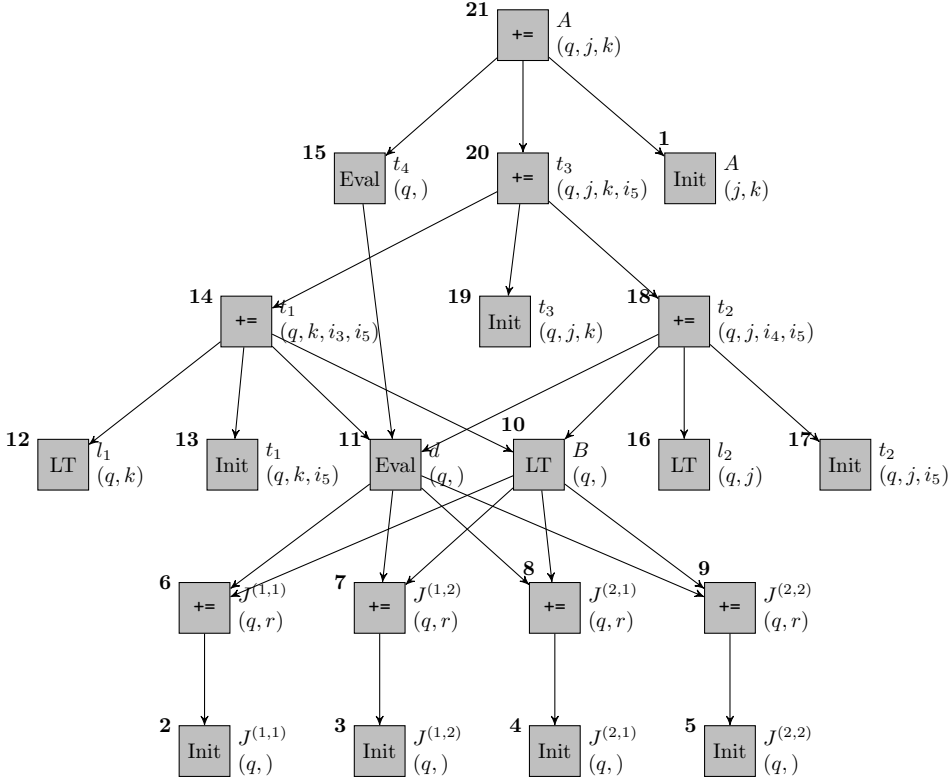}

    \node[right,align=left] at (J11_qi.east) (J11_qi_fi) {$J^{(1,1)}$\\$(q,)$};
    \node[right,align=left] at (J12_qi.east) (J12_qi_fi) {$J^{(1,2)}$\\$(q,)$};
    \node[right,align=left] at (J21_qi.east) (J21_qi_fi) {$J^{(2,1)}$\\$(q,)$};
    \node[right,align=left] at (J22_qi.east) (J22_qi_fi) {$J^{(2,2)}$\\$(q,)$};
    \node[right,align=left] at (J11_qa.east) (J11_qa_fi) {$J^{(1,1)}$\\$(q,r)$};
    \node[right,align=left] at (J12_qa.east) (J12_qa_fi) {$J^{(1,2)}$\\$(q,r)$};
    \node[right,align=left] at (J21_qa.east) (J21_qa_fi) {$J^{(2,1)}$\\$(q,r)$};
    \node[right,align=left] at (J22_qa.east) (J22_qa_fi) {$J^{(2,2)}$\\$(q,r)$};

    \node[right,align=left] at (B_q.east) (B_q_fi) {$B$\\$(q,)$};
    \node[right,align=left] at (d_q.east) (d_q_fi) {$d$\\$(q,)$};
    \node[right,align=left] at (t4_q.east) (t4_q_fi) {$t_4$\\$(q,)$};

    \node[right,align=left] at (l1_qk.east) (l1_qk_fi) {$l_1$\\$(q,k)$};
    \node[right,align=left] at (l2_qj.east) (l2_qj_fi) {$l_2$\\$(q,j)$};
    \node[right,align=left] at (t1_qk5i.east) (t1_qk5i_fi) {$t_1$\\$(q,k,i_5)$};
    \node[right,align=left] at (t2_qj5i.east) (t2_qj5i_fi) {$t_2$\\$(q,j,i_5)$};
    \node[right,align=left] at (t3_qjki.east) (t3_qjki_fi) {$t_3$\\$(q,j,k)$};
    \node[right,align=left] at (t1_qk5a.east) (t1_qk5a_fi) {$t_1$\\$(q,k,i_3,i_5)$};
    \node[right,align=left] at (t2_qj5a.east) (t2_qj5a_fi) {$t_2$\\$(q,j,i_4,i_5)$};
    \node[right,align=left] at (t3_qjka.east) (t3_qjka_fi) {$t_3$\\$(q,j,k,i_5)$};

    \node[right,align=left] at (R_jki.east) (R_jki_fi) {$A$\\$(j,k)$};
    \node[right,align=left] at (R_jka.east) (R_jka_fi) {$A$\\$(q,j,k)$};

    \node[above] at (R_jki.north west) (e1) {\textbf{1}};
    \node[left] at (J11_qi.north west) (e2) {\textbf{2}};
    \node[left] at (J12_qi.north west) (e3) {\textbf{3}};
    \node[left] at (J21_qi.north west) (e4) {\textbf{4}};
    \node[left] at (J22_qi.north west) (e5) {\textbf{5}};
    \node[left] at (J11_qa.north west) (e10) {\textbf{6}};
    \node[left] at (J12_qa.north west) (e11) {\textbf{7}};
    \node[above] at (J21_qa.north west) (e12) {\textbf{8}};
    \node[above] at (J22_qa.north west) (e13) {\textbf{9}};
    \node[above] at (B_q.north west) (e14) {\textbf{10}};
    \node[left] at (d_q.north west) (e15) {\textbf{11}};
    \node[left] at (l1_qk.north west) (e16) {\textbf{12}};
    \node[left] at (t1_qk5i.north west) (e17) {\textbf{13}};
    \node[left] at (t1_qk5a.north west) (e18) {\textbf{14}};
    \node[left] at (t4_q.north west) (e19) {\textbf{15}};
    \node[left] at (l2_qj.north west) (e20) {\textbf{16}};
    \node[left] at (t2_qj5i.north west) (e21) {\textbf{17}};
    \node[left] at (t2_qj5a.north west) (e22) {\textbf{18}};
    \node[left] at (t3_qjki.north west) (e23) {\textbf{19}};
    \node[left] at (t3_qjka.north west) (e24) {\textbf{20}};
    \node[left] at (R_jka.north west) (e25) {\textbf{21}};
  \end{tikzpicture}
  \caption{\emph{Dependency graph} of the loop nests of
    \cref{alg:unfused}.  At the right of each node are the name of
    the \emph{temporary variable} and the ordered loop indices of the
    loop nest.  The bold numbers at the upper left of each node
    represent a valid ordering of these operations.}
  \label{fig:nest_graph}
\end{figure}

\Cref{fig:nest_graph} shows the \emph{dependency graph} of the
loop nests in \cref{alg:unfused}, along with the solution our
algorithm yields.  The ordering precisely corresponds to what we have
seen in \cref{alg:reordered}.

\subsection{Related work}
\label{sec:tensor_work}

Since efficient implementation of tensor algebra expressions is not
specific to the finite element method, the related work is broad in
scope.  Various tensor algebra libraries and domain-specific languages
\cite{Baumgartner2005,Epifanovsky2013,Sanders2010,Solomonik2013} have
been developed for quantum chemistry computations.  However, while the
\emph{working set size}\footnote{The amount of memory required to
  carry out a sequence of operations.} of a finite element local
assembly kernel rarely exceeds a megabyte, the tensors in quantum
chemistry computations are typically huge, even reaching petabytes in
size.  Consequently, the optimized implementation of individual tensor
operations on distributed-memory parallel computers is usually the
major concern, as is the case with SIAL \cite{Sanders2010}, libtensor
\cite{Epifanovsky2013}, and CTF \cite{Solomonik2013}.  Similarly,
TensorFlow \cite{Abadi2016} serves to instrument tensor computations
for machine learning purposes on large-scale heterogeneous computing
systems, so most of their challenges do not overlap with ours.  The
\emph{Tensor Contraction Engine} (TCE) \cite{Baumgartner2005},
however, also considers optimizations on the whole tensor algebra
expression, which are applicable to our case as well.  Some of these
ideas were also applied to do efficient tensor contractions on GPUs
\cite{Nelson2015}.

Some other projects also feature tensor expressions as intermediate
representations.  SPL \cite{Xiong2001} is a component of the SPIRAL
system \cite{Puschel2005}, and its language is quite similar to GEM,
but tuned for signal processing problems such as the fast Fourier
transform (FFT).  $ \Sigma $-SPL \cite{Franchetti2005} is an extension
that features index mappings similar to the \texttt{FlexiblyIndexed}
nodes in GEM.  EIN \cite{Chiw2016} is another tensor language, and the
intermediate representations of Diderot \cite{Chiw2012,Kindlmann2016}.
It uses a \emph{static single assignment} (SSA) form, while GEM is a
simple expression language that cannot express control flow.  Although
this choice limits the kind of kernels GEM can represent, GEM is also
easier to reason about.

\emph{Algebraic transformations} on tensor algebra expressions can
produce equivalent formulations that require many fewer operations to
compute.  Lam et~al.\@ \cite{Lam1996,Lam1997} first studied the
problem of \emph{single-term optimization}, also known as
\emph{strength reduction}: the operation-minimal evaluation of tensor
contractions over the product of several tensors with various indices.
They prove that the problem is NP-complete, but they also provide a
pruning search strategy that is efficient in practice, since a single
term typically consists of only a limited number of tensors.
Kschischang et~al.\@ \cite{Kschischang2001} show that such a single
term can be characterised by a \emph{factor graph}.  If the factor
graph is \emph{cycle-free} then it straightforwardly yields the unique
optimal factorisation.  Otherwise different factorisations can be
produced by various transformations that break the cycles.
Considering the sum of single terms, Hartono et~al.\@
\cite{Hartono2005} achieved further improvements over single-term
optimization by considering various factorisations of the whole
expression.  Since the space of different factorisations is huge, they
relied on various search techniques.  Exploiting \emph{common
  subexpression elimination} during the search resulted in further
gains \cite{Hartono2006}.  TSFC primarily relies on COFFEE to provide
operation minimization (see \cref{sec:coffee}).

Although rearranging compound expressions may drastically reduce the
required computational work in some cases, the efficient
implementation of primitive tensor operations is necessary for the
effective use of hardware.  Traditionally, BLAS (Basic Linear Algebra
Subroutines) implementations provide optimized dense linear algebra
operations; Springer and Bientinesi \cite{Springer2016} present
efficient strategies for generalising to tensor-tensor multiplication.
However, these routines are optimized for large tensors and matrices.
Eigen \cite{Guennebaud2010} is a templated, header-only C++ library
that provides efficient linear algebra routines for small tensors.
M\"{u}thing et~al.\@ \cite{Muething2017} use libxsmm \cite{Heinecke2016}
in their high-order discontinuous Galerkin code, which contains
heavily optimized BLAS-like routines for small matrices.  Kjolstad
et~al.\@ \cite{Kjolstad2017} present the \emph{Tensor Algebra
  Compiler} (\texttt{taco}), which generates efficient code for
primitive tensor operations, especially for various sparse array
formats.  While finite element local assembly mostly involves dense
tensor operations, Kirby \cite{Kirby2017a} describes a theory of
reference transformation that requires multiplications with sparse
matrices for finite elements with stronger than $ C^0 $ continuity.

Since tensors in quantum chemistry computations are huge, optimizing
the memory requirement is of great importance so that computations can
fit in the main memory.  While this is not an issue for finite element
local assembly, fitting the working set into the cache can have an
important performance impact.  The evaluation of tensor expressions in
a scalar programming language such as C requires the introduction of
\emph{array temporaries} which may be indexed.  The order in which
these temporaries are produced and consumed affects the working set
size of the kernel.  Lam et~al.\@ \cite{Lam1999a,Lam2011} give an
algorithm for the evaluation order of nodes of an expression tree that
minimises the maximal total size of live temporaries.  Their work can
be understood as a generalisation of the Sethi-Ullman
algorithm~\cite{Sethi1970} where nodes are not assumed to have the
same size, instead each node can have an arbitrary size.  Since tensor
valued temporaries are both produced and consumed by loops, \emph{loop
  fusion} can eliminate dimensions of these temporaries, thus reducing
the required memory, as we will demonstrate in \cref{sec:codegen}.
Lam et~al.\@ \cite{Lam1999} formalise the legality of loop fusions
over expression trees and their effect on the memory requirement. They
give an algorithm for finding the memory-optimal set of loop fusions
based on dynamic programming and pruning, with an exponential
complexity in the number of index variables.  Allam et~al.\@
\cite{Allam2006} propose an alternative algorithm that reduces this
problem to an \emph{integer linear programming} formulation.  Cociorva
et~al.\@ \cite{Cociorva2002} studied space-time trade-off optimization
in this context, by allowing some recomputations for further loop
fusion opportunities.  In TSFC we have implemented a fast algorithm
that may produce suboptimal results
(\cref{sec:codegen,sec:scheduling}), rather than striving for
optimality at the cost of high algorithmic complexity
\cite{Lam1999,Allam2006}.

\section{Optimizations}
\label{sec:opt}

This section discusses various optimizations in TSFC which are not
essential for the understanding of the overall form compilation
process, and whose proper effects are often only understood in
composition with other optimizations at various points of the form
compilation pipeline.

\subsection{Constant folding for zero tensors}
\label{sec:zero_folding}

\emph{Constant folding} means replacing computations whose all
operands are known at compile time with the result of the computation.
This is a standard technique in modern optimizing compilers.  TSFC
implements a limited version of it at the GEM level: only computations
involving \emph{zero tensors} are folded.  This limited version can
still yield significant benefits since the application of
$ 0 \cdot X = 0 $ can eliminate an arbitrarily complex expression
$ X $.  Zero tensors can arise, for example, as tabulation matrices
for higher derivatives than the degree of the finite element.

\subsection{Cellwise constant quantities}
\label{sec:cellwise_const}

We saw in \cref{sec:fem} that the evaluation of arguments and
coefficients involves a tabulation matrix $ M_{q,r} $ that contains
the value of each basis function at each quadrature point.  When an
argument or a coefficient, or more commonly a derivative of it, is
cellwise constant, then evaluation need not happen for every
quadrature point.  So instead of the matrix $ M_{q,r} $ we create a
vector $ M_r $ which does not have the quadrature index as a free
index.  This change in combination with loop hoisting (see
\cref{sec:codegen}) will hoist the evaluation of all cellwise
constant subexpressions out of the quadrature loop.

\subsection{Fast Jacobian evaluation for affine cells}
\label{sec:fast_jacobian}

The Jacobian of the coordinate transformation appears in virtually
every form after preprocessing, even the simplest ones.  In
those cases an inefficient Jacobian evaluation may likely become the
dominant cost of assembly.  Consequently, FEniCS has, and Firedrake
used to have, efficient hand-written code snippets for Jacobian
evaluation on affine simplices.  As described in \cref{sec:fem}, we
now implement Jacobian evaluation as coefficient evaluation.  This
allows us to handle arbitrary order geometries with a single code
path, but it also implies that each entry of the Jacobian is evaluated
as a matrix-vector product.  For affine cells, the Jacobian is cellwise
constant, so this reduces to a dot product of two vectors
(\cref{sec:cellwise_const}).  The code snippets evaluate each
Jacobian entry with a single subtraction.

Since both approaches are correct and thus yield identical results,
it follows immediately that the \emph{tabulation vector} must have
precisely two nonzero entries which are $ 1 $ and $ -1 $.  Therefore,
we introduced another optimization rule which applies to all
coefficient evaluation: if the quantity is cellwise constant and the
tabulation vector has at most two nonzero entries, then we expand the
dot product that evaluates the coefficient.  This, of course, composes
with constant folding (\cref{sec:zero_folding}), so the
multiplications with zeros vanish, and we obtain a formulation that is
performance equivalent to the code snippets.

\subsection{Unrolling short \texttt{IndexSum} loops}
\label{sec:unroll_indexsum}

In most cases, unrolling short loops, which are typically geometric
loops, improves performance.  When this optimization is applied, the
generated code for the Laplace operator becomes as in
\cref{alg:unroll_indexsum}.  It also negates the code duplication
around Jacobian inverses that was caused by the \emph{beta expansion}
step (\cref{sec:component_tensor}).  This optimization is enabled by
default in TSFC.  However, as demonstrated, it is completely optional,
and can be delayed to run after those optimizations that benefit from
the tensor structure of the source expression.

\begin{algorithm}
  \begin{multicols}{2}
  \begin{algorithmic}[1]
    \ForAll {$ j, k $}
    \State $ A[j, k] \leftarrow 0 $
    \EndFor
    \ForAll {$ q $}
    \State $ J^{(1,1)} \leftarrow 0 $
    \State $ J^{(1,2)} \leftarrow 0 $
    \State $ J^{(2,1)} \leftarrow 0 $
    \State $ J^{(2,2)} \leftarrow 0 $
    \ForAll {$ r $}
    \State $ J^{(1,1)} $ += $ C_{q,r}^{(1,1)} * c_r $
    \State $ J^{(1,2)} $ += $ C_{q,r}^{(1,2)} * c_r $
    \State $ J^{(2,1)} $ += $ C_{q,r}^{(2,1)} * c_r $
    \State $ J^{(2,2)} $ += $ C_{q,r}^{(2,2)} * c_r $
    \EndFor
    \State $ d = J^{(1,1)} * J^{(2,2)} - J^{(1,2)} * J^{(2,1)} $
    \State $ K^{(1,1)} = J^{(2,2)} \div d $
    \State $ K^{(1,2)} = -J^{(1,2)} \div d $
    \State $ K^{(2,1)} = -J^{(2,1)} \div d $
    \State $ K^{(2,2)} = J^{(1,1)} \div d $
    \ForAll {$ k $}
    \State $ t_1[k] = K_{11} E^{(1)}_{q,k} + K_{21} E^{(2)}_{q,k} $
    \State $ t_2[k] = K_{12} E^{(1)}_{q,k} + K_{22} E^{(2)}_{q,k} $
    \EndFor
    \State $ t_5 = w_q * \mathrm{abs}(d) $
    \ForAll {$ j $}
    \State $ t_3 = K_{11} E^{(1)}_{q,j} + K_{21} E^{(2)}_{q,j} $
    \State $ t_4 = K_{12} E^{(1)}_{q,j} + K_{22} E^{(2)}_{q,j} $
    {\small
    \ForAll {$ k $}
    \State \hspace{-1.4em} $ A[j, k] $ += $ t_5 * (t_1[k] * t_3 + t_2[k] * t_4) $
    \EndFor
    }
    \EndFor
    \EndFor
  \end{algorithmic}
  \end{multicols}
  \caption{Generated code when unrolling applied, equivalent to
    \cref{alg:fused}.}
  \label{alg:unroll_indexsum}
\end{algorithm}

\subsection{COFFEE integration}
\label{sec:coffee}

Although we did not emphasize this, TSFC does not generate C code
directly, but an \emph{abstract syntax tree} (AST) for C, so that
COFFEE can further optimize the kernel.  Since this is just a
different representation of C code, the discussions in
\cref{sec:gem2c} are still completely valid.  COFFEE performs both
low-level optimizations, such as alignment and padding to facilitate
vectorization by the host C compiler, and high-level optimizations to
reduce the operation count.  The latter require restructuring the
loops and expressions.  In essence, COFFEE seeks to determine the
optimal trade-off amongst common subexpressions elimination,
loop-invariant code motion and factorization.  For this, it analyzes a
transformation space which also includes the FFC tensor representation
and many of the factorizations performed by UFLACS.
For further
details about COFFEE and for comparative performance evaluations, see
Luporini, Ham, and Kelly \cite{Luporini2016}.

\section{Experimental evaluation}
\label{sec:exp}

We experimentally evaluate and compare form compilation times with
the following form compilers:
\begin{itemize}
\item FFC quadrature representation (supports affine cells only)
\item Modified version of FFC quadrature representation, developed for
  Firedrake to support non-affine cells.
\item UFLACS
\item TSFC
\end{itemize}
We disabled optional optimizations, and also included in the form
compilation times the time to compile the generated C or C++ code,
using GCC 5.4.0 as C and C++ compiler at \texttt{-O2} optimization level.
FFC tensor representation is not included since its limitations are
well known \cite{Oelgaard2010} (\cref{sec:fc}), and it fails to
compile our more difficult test problems due to missing features.

Our test problems consist of the left-hand sides of four elliptic PDEs
of increasing complexity: the Helmholtz equation, linear elasticity, a
simple hyperelastic model, and the Holzapfel-Ogden model as described
in \cite{Balaban2016}.  A full description is provided in supplement
\cref{sec:test_cases}.

\Cref{fig:data} presents violin plots of the form compilation times,
each made from 18 samples.  The measurements were run on an
Intel\textsuperscript{\textregistered}
Xeon\textsuperscript{\textregistered} E5--2620 (Sandy Bridge) 2.00 GHz
CPU.  Due to the nature of development with Firedrake or FEniCS, the
form compiler is frequently invoked in an interactive manner.
Consequently, form compilation shall preferably take no more time than
the fraction of a second; waiting for more than a couple of seconds is
undesirable in a natural workflow.

\begin{figure}
  \centering
  \input{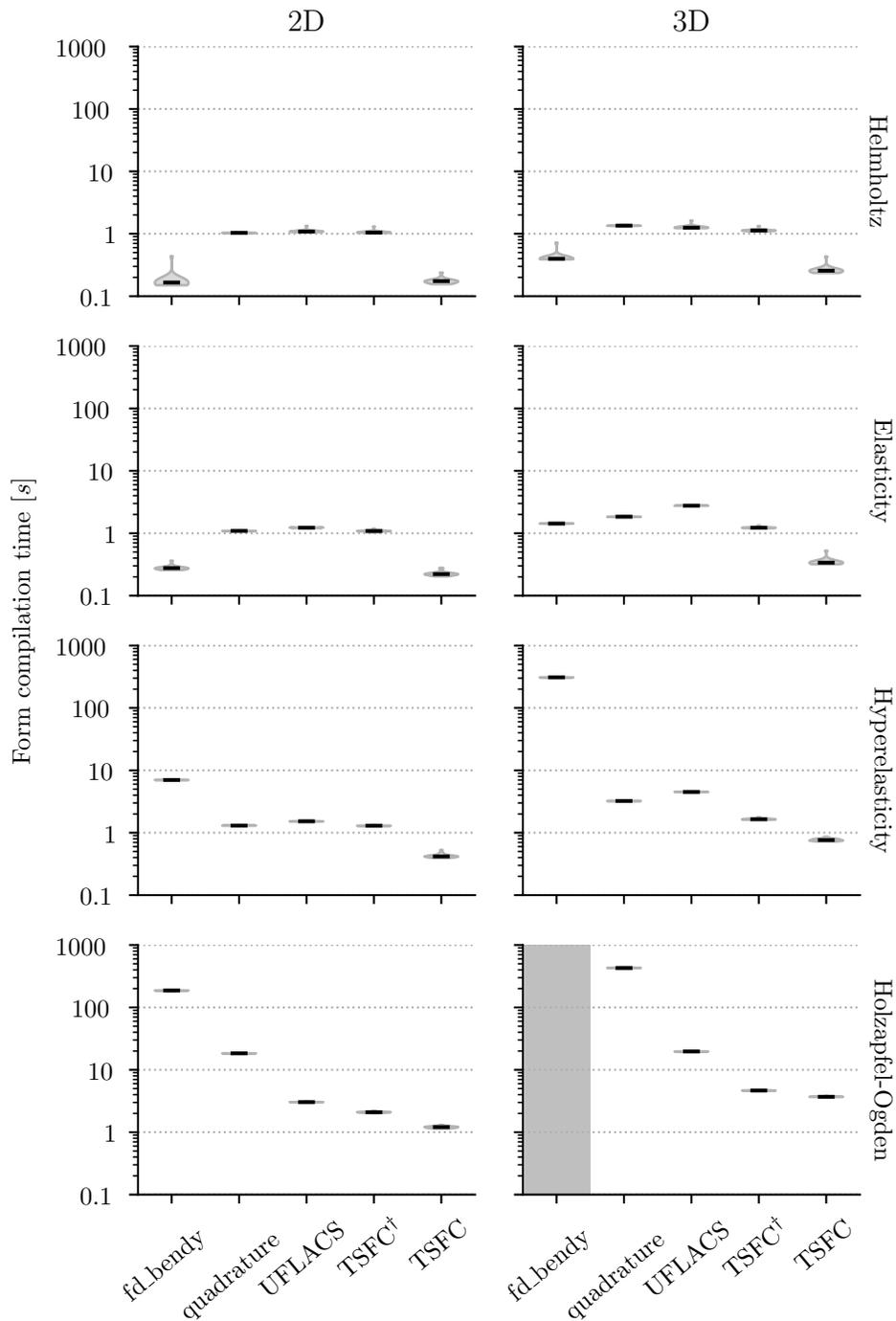}
  \caption{Form compilation time for forms of increasing complexity
    with the following form compilers: modified version of FFC
    quadrature representation which supports non-affine cells
    (fd\_bendy), upstream FFC quadrature, UFLACS, TSFC as an FFC
    representation (TSFC$^\dagger$), and standalone TSFC.  3D
    Holzapfel-Ogden data for fd\_bendy are missing because compilation
    ran out of memory after more than 6 hours on a machine with 64GB
    of memory.}
  \label{fig:data}
\end{figure}

To make the measurements as fair as possible, we configured FFC
representations to skip element compilation, and for the C++
compilation we only kept the relevant \texttt{tabulate\_tensor} method
from the generated UFC class.  However, since we did not want to
change the generated code, we had to include the relevant header
files.  FFC representations generate code that use functions from the
standard C++ header \texttt{<algorithm>}, the compilation of which
takes about 0.9 seconds on the test machine.  Form compilers targeting
Firedrake (fd\_{}bendy and TSFC) generate C code, and the inclusion of
relevant header files induces much less overhead.  This effect
dominates the simplest test case, the Helmholtz equation.  For further
fairness, we also included TSFC as an FFC representation (shown as
TSFC$^\dagger$), including the same set of header files as for the
other FFC representations.

All form compilers compile the Helmholtz equation quickly and perform
tolerably on the elastic model. The modified version of FFC quadrature
with non-affine support (fd\_bendy) demonstrates excessive compilation
times on the simple hyperelastic model already.  While non-affine
support adds a huge toll to compile-time performance, even the base
version of FFC quadrature becomes unmanageable as form complexity
increases, as the data for Holzapfel-Ogden model demonstrate.  UFLACS
and TSFC, however, have tolerable compile-time performance in all
cases, TSFC being somewhat faster, but typically within the same order
of magnitude.

The overall compilation time of the most difficult case
(Holzapfel-Ogden 3D) with TSFC can further be reduced from 3.7 to 2.2
seconds by disabling the optimization described in
\cref{sec:unroll_indexsum}.  This configuration, however, did not
yield significant benefits for the other test cases as long as C
compilation times are also included.

For reproducibility, we cite archives of the exact software versions
that were used to produce these results.  The modified version of FFC
with support for non-affine geometries requires custom branches of
FFC~\cite{anders_logg_2017_573237} and
UFL~\cite{martin_sandve_alnaes_2017_573236} which require older
versions of COFFEE~\cite{fabio_luporini_2017_573235},
FIAT~\cite{miklos_homolya_2017_573238}, and
Instant~\cite{miklos_homolya_2017_573255}.  For all other form
compilers and representations we used recent versions:
COFFEE~\cite{fabio_luporini_2017_573267},
dijitso~\cite{martin_sandve_alnaes_2017_573287},
FFC~\cite{miklos_homolya_2017_573270},
FIAT~\cite{miklos_homolya_2017_573269},
FInAT~\cite{david_a_ham_2017_573266},
TSFC~\cite{miklos_homolya_2017_573271}, and
UFL~\cite{martin_sandve_alnaes_2017_573268}.
The experimentation framework used to generate the violin plots is
available in \cite{miklos_homolya_2017_573371}.

\section{Conclusion}
\label{sec:conclusion}

The experimental data demonstrate that the efficient traversal of
expressions keeps the compilation time with TSFC acceptably low even
for highly complicated forms.  This was a crucial requirement for
properly supporting non-simplicial cells as well as higher-order
geometries in Firedrake.

A comprehensive analysis of run-time performance (of the generated
code) is beyond the scope of this paper.  For such an analysis, we
refer readers to Luporini, Ham, and Kelly \cite{Luporini2016} which
compares the representations of FFC, UFLACS, and an earlier and a
recent version of COFFEE.  TSFC is compatible with COFFEE's
optimizations.

\section*{Acknowledgements}

The authors would like to thank Andrew T.\@ T.\@ McRae for reviewing
the manuscript of this paper.  The integration of TSFC as an FFC
representation was undertaken by Jan Blechta.  TSFC relies on several
recent enhancements to UFL by Martin S.\@ Aln\ae{}s and Andrew T.\@
T.\@ McRae.

\bibliographystyle{siamplain}
\bibliography{references,zenodo}

\end{document}